\documentclass[twocolumn]{aastex63}

\graphicspath{{./}{figures/}}

\received{March 29, 2021}
\accepted{Oct 25, 2021}

\shorttitle{Clearing the hurdle: The mass of GC systems as a function of host galaxy mass}
\shortauthors{Eadie, Harris, \& Springford}

\usepackage{longtable}
\usepackage{threeparttable}
\usepackage{bm}
\usepackage{mathrsfs}
\usepackage{lineno}
\usepackage{amsmath}

\usepackage{color}
\usepackage{fancyvrb}

\DefineVerbatimEnvironment{Highlighting}{Verbatim}{commandchars=\\\{\}}
\usepackage{framed}
\definecolor{shadecolor}{RGB}{248,248,248}
\newenvironment{Shaded}{\begin{snugshade}}{\end{snugshade}}

\newcommand{\AttributeTok}[1]{\textcolor[rgb]{0.77,0.63,0.00}{#1}}

\newcommand{\CommentTok}[1]{\textcolor[rgb]{0.56,0.35,0.01}{\textit{#1}}}

\newcommand{\FunctionTok}[1]{\textcolor[rgb]{0.00,0.00,0.00}{#1}}

\newcommand{\NormalTok}[1]{#1}

\newcommand{\OtherTok}[1]{\textcolor[rgb]{0.56,0.35,0.01}{#1}}

\newcommand{\SpecialCharTok}[1]{\textcolor[rgb]{0.00,0.00,0.00}{#1}}

\newcommand{\StringTok}[1]{\textcolor[rgb]{0.31,0.60,0.02}{#1}}

\newcommand{\uat}[2]{\href{http://astrothesaurus.org/uat/#2}{#1 (#2)}}

\begin{document}

\title{\textbf{Clearing the hurdle: The mass of globular cluster systems as a function of host galaxy mass}}

\correspondingauthor{Gwendolyn M. Eadie}
\email{gwen.eadie@utoronto.ca}
\author[0000-0003-3734-8177]{Gwendolyn M. Eadie}
\affil{David A. Dunlap Department of Astronomy \& Astrophysics, University of Toronto, Toronto, ON M5S 3H4, Canada}
\affil{Department of Statistical Sciences, University of Toronto, Toronto, ON M5S 3G3, Canada}

\author[0000-0001-8762-5772]{William E. Harris}
\email{harris@physics.mcmaster.ca}
\affil{Department of Physics and Astronomy, McMaster University, Hamilton, ON L8S 4M1, Canada}

\author[0000-0001-7179-9049]{Aaron Springford}
\email{aaron.springford@cytel.com}
\affil{Cytel, 1 University Avenue, 3rd Floor, Toronto, ON M5J 2P1, Canada}

\begin{abstract}
Current observational evidence suggests that all large galaxies contain globular clusters (GCs), while the smallest galaxies do not. Over what galaxy mass range does the transition from GCs to no GCs occur? We investigate this question using galaxies in the Local Group, nearby dwarf galaxies, and galaxies in the Virgo Cluster Survey. We consider four types of statistical models: (1) logistic regression to model the probability that a galaxy of stellar mass $M_*$ has any number of GCs; (2) Poisson regression to model the number of GCs versus $M_*$, (3) linear regression to model the relation between GC system mass ($\log{M_{gcs}}$) and host galaxy mass ($\log{M_{\star}}$), and (4) a \textit{Bayesian lognormal hurdle model} of the GC system mass as a function of galaxy stellar mass for the entire data sample. From the logistic regression, we find that the 50\% probability point for a galaxy to contain GCs is $M_{*}=10^{6.8}M_{\odot}$. From post-fit diagnostics, we find that Poisson regression is an inappropriate description of the data. Ultimately, we find that the Bayesian lognormal hurdle model, which is able to  describe how the mass of the GC system varies with $M_*$ even in the presence of many galaxies with no GCs, is the most appropriate model  over the range of our data. In an Appendix, we also present photometry for the little-known GC in the Local Group dwarf Ursa Major II.
\end{abstract}



\section{Introduction}\label{sec:intro}

Globular clusters (GCs) --- old massive star clusters formed
in the early stages of galaxy assembly --- are generally found in all galaxies except the smallest dwarfs \citep{harris2010,forbes+2018,beasley2020}. For galaxies with GC populations, empirical evidence suggests that the total mass in GCs ($M_{gcs}$)  is nearly proportional to the host galaxy's total halo mass ($M_h$) rather than its total stellar mass $M_{\star}$ \citep[e.g.][]{blakeslee1997,spitler_forbes2009,georgiev+2010,hudson+2014,
harris+2017}. This observation has stimulated a growing body of theoretical modelling of GC formation during galaxy hierarchical growth at high redshift \citep[e.g.][]{kravtsov_gnedin2005,boylan-kolchin2017,pfeffer+2018,choksi+2018,li_gnedin2019, choksi_gnedin2019,bastian+2020}.

The $M_{gcs} - M_h$ relation has focused primarily on luminous galaxies, and little is known about the relation for the smallest dwarfs.  For these tiny systems, it is extremely difficult to determine their halo masses empirically \citep{forbes+2018}, and even the model fits that rely on abundance matching and other methods \citep[e.g.][]{behroozi+2013,moster+2010,hudson+2015,read+2017}
require extrapolation to obtain any kind of $M_h$ estimate for the ultra-faint dwarfs. If the $M_{gcs}-M_h$ relation extended downward in a truly linear fashion, it would, for example, mean that an ultra-faint dwarf galaxy consisting of only a dark matter potential well with almost no stars should still hold a GC population. Observationally such an interpretation may be ruled out
\citep[e.g.][and see below]{amorisco+2018,forbes+2018,forbes+2020}, whereas recent theory favors extension of the standard relation downward, though not
linearly \citep[see][]{el-badry+2019,choksi_gnedin2019,bastian+2020}.

New observational campaigns that search for ultra-faint galaxies with GC populations could  change our current understanding. In the meantime, however, a better understanding of the transition region between galaxies with GCs and those without could elucidate both the early formation and later evolution of star clusters in small galaxies. Thus, we would like to answer the question: what is the \textit{transition mass region} for galaxies over which galaxies go from not having GCs to having them?

In this paper, we investigate empirical, statistical models that might illuminate this transition region and that could describe the observed relationship between $M_{\star}$, GC systems, and galaxy type. We look for a transition mass region by using models for both binary data and continuous data. Response variables may be (a) whether or not a galaxy has a GC population, (b)  the total number of GCs $N_{GC}$ in a galaxy, (c) the total mass of the GC system $M_{gcs}$, or (d) some combination of the previous three. Throughout, we use the the galaxy stellar mass  $M_{\star}$ as the covariate or predictor.

Our data include galaxies from the Local Group, a sample of dwarf galaxies, and the Virgo Cluster Survey, in an attempt to construct a sample over the full necessary range of galaxy masses. For the majority of our sample, no independent measure of halo mass $M_h$ exists, so of necessity we restrict our discussion to the total stellar mass $M_{\star}$.  

Our paper is structured as follows:
\begin{itemize}
    \item In Section~\ref{sec:data}, we define our observational sample of data. 
    
    \item In Section~\ref{sec:EDA}, we perform some simple exploratory data analysis, looking at the relationships observed between $\log{M_{\star}}$ and various response variables, such as whether or not a galaxy has GCs, the number of GCs $N_{gc}$, and the mass in the GC population $M_{gcs}$. This motivates the next four sections where we use empirical models to describe the observed relationships.
    
    \item In Section~\ref{sec:LR}, we use \emph{logistic regression} to model the relationship between galactic stellar mass and whether or not a galaxy has GCs. First, we review the basics of logistic regression (Section~\ref{sec:logisticregression}), and then present our results for the LG and the entire sample.
     
    \item In Section~\ref{sec:poisson}, we apply\emph{Poisson regression} to model $N_{gcs}$  versus $M_{\star}$, and assess the model's ability to describe the entire data set at all mass ranges.
     
     \item In Section \ref{sec:mass}, we explore the observed trend of $M_{gcs}$ versus $M_{\star}$ for galaxies with GCs through the lens of a linear model. 
     
     \item In Section~\ref{sec:hurdle} we combine the use of logistic regression with linear regression via a \textit{Bayesian lognormal hurdle model}, introducing the basics of the procedure and applying it to our entire data set.
     
     \item Lastly, in Section~\ref{sec:conclusions} we make comparisons with 
     recent related studies and discuss possible physical interpretations of our findings. A summary is found in Section \ref{sec:summary}.
\end{itemize}

In what follows, all masses implicitly have units of Solar mass $M_{\odot}$ unless otherwise stated. To ease notation, any logarithmic quantities of base ten (log$_{10} x$) and base $e$ are written simply as $\log{x}$ and $\ln x$ respectively. For example, instead of  $\log_{10}{\frac{M_{\star}}{M_{\odot}}}$, we simply write $\log{M_{\star}}$.

\section{The Observational Sample}\label{sec:data}

We are interested in whether or not the mere presence of GCs is related to galaxy mass, and if there is a definable transition mass region between galaxies with or without GCs. To infer this relationship, we need a sample of galaxies with measurements of total stellar mass $M_{\star}$, the GC system mass $M_{gcs}$, and an indicator of whether or not the galaxy has a GC population. 

We assemble a data set from three sources: the Local Group, \citet{georgiev+2010} (nearby dwarfs), and the HST/ACS Virgo Cluster Survey (VCS) \citep{cote+2004}. Where necessary, the galaxy luminosities $M_V^T$ in all three of our datasets are converted to total stellar mass $M_{\star}$ with the mass-to-light ratio 
\begin{equation}
    {\rm log} (M/L_V) = -0.778 + 1.305 (B-V)_0
\end{equation}
from \citet{bell+2003} and with the assumption of a Chabrier IMF.

\subsection{Local Group Sample}\label{sec:LG}

The first sample consists of the Local Group (LG) galaxies. Compared with all other environments, the LG sample has the advantages of the best available \emph{homogeneity, depth, and completeness:} all types of galaxies except giant early-type galaxies (ETGs) are represented; the census for the LG is highly complete down to extremely faint galaxy luminosity levels;  and searches for GCs have been carried out for all of them. Moreover, we note that the largest galaxy members all have GCs while the smallest ones do not \citep[e.g.][]{huang_koposov2020}. The Local Group member lists are taken from \cite{simon2019} and \citet{drlica-wagner+2020}, to which we add the Milky Way satellite Gaia/Enceladus as discussed by \citet{forbes2020}, and the star cluster populations in IC10 by \citet{lim_lee2015}\footnote{We specifically exclude the anomalous case of M32, the cE satellite of M31.  The number of GCs belonging to M32 is unknown and may be zero, generally thought to be a consequence of tidal stripping of its envelope and severe dynamical removal of any inner clusters it may once have had; e.g., \citet{choi+2002,brockamp+2014,d'souza_bell2018}.}. We also add the small star cluster in the dwarf Ursa Major II (see Appendix~\ref{app:photometry}).  Thus, in our LG sample, there are a total of 100 galaxies, with 20 of these having GCs (Table~\ref{tab:summarydata}).  The numbers of GCs ($N_{GC}$) in each LG member and their total mass ($M_{gcs}$) are taken from \citet{harris+2013}, \citet{forbes2020}, and \citet{forbes+2018}.
Although the presence of a few very faint and undiscovered GCs in the LG dwarfs cannot yet be
ruled out (see the Ursa Major II GC in Appendix~\ref{app:photometry} as an example), the LG still gives us the most complete sample over the
largest range of galaxy luminosities.

\subsection{Nearby Dwarf Sample}\label{sec:dwarfs}

The second sample is from the work of \citet{georgiev+2010}, who studied the GC populations in  nearby dwarf galaxies that are isolated or in small-group environments not including the Local Group.  The lower limit of their dwarf galaxy magnitudes is near $M_V(lim) \simeq -10$, or log $M_{\star} \simeq 6.3$.  GC system masses were drawn directly from their  paper.  Notably, \citet{georgiev+2010} show that the fraction $f$ of galaxies with GCs, plotted versus bins in luminosity (see their Figure 4), displays a gradual falloff of $f$ toward low luminosity. The 50\% point ($f=0.5$) is near $M_V \simeq -11.5$, which  corresponds to $\log{M_{\star}} \simeq 6.8 \pm 0.2$.

\subsection{Virgo Cluster Survey Sample}\label{sec:VCS}

Both the LG and the \citet{georgiev+2010} dwarf samples have the limitation that they include few higher-luminosity galaxies; the former is not populous enough to contain numerous massive galaxies, and the latter deliberately includes only dwarfs. Thus, for our third sample we add in 93 systems from the HST/ACS Virgo Cluster Survey (VCS, where the data are taken from \citet{peng+2008,jordan+2009}). The lowest-luminosity systems in the VCS are $\sim 4 \times 10^8 M_{\odot}$, but the survey has the advantages that all members are at a common distance. The sample is also homogeneous and thorough in identifying the GC populations in its target list, with little residual contamination from non-GC objects. 
Here we exclude galaxies with $M_* > 10^{11} M_{\odot}$, since we are interested in the lower-mass regime.

GC systems have also been systematically investigated in other surveys of low-luminosity galaxies, for example within the Coma cluster of galaxies \citep{vandokkum+2017,lim+2018,amorisco+2018}, Fornax \citep{prole+2019}, Virgo \citep{lim+2020}, and smaller groups \citep{somalwar+2020}, primarily concentrating on their Ultra-Diffuse Galaxies (UDGs). These samples cover a range of masses we are interested in, but the GC numbers are not determined by direct identification of GCs. The estimated GC numbers in these cases are residuals after background subtraction, and in most cases, the uncertainty in $N_{GC}$ is larger than $N_{GC}$ itself. Because of this large uncertainty, we exclude these lists in the current analysis. However, exactly how to incorporate these and their uncertainties into the empirical models used here could be an interesting avenue of future work.

\section{Exploratory Data Analysis and a Transition Mass Region}\label{sec:EDA}

Table~\ref{tab:summarydata} is a contingency table for our entire sample of 232 galaxies (LG, nearby dwarfs, and VCS), showing how many galaxies in each sample have or do not have GCs.

\begin{table}[ht]
\centering
\caption{Contingency table for the entire sample; the number of galaxies within each sample that do not have (0) and have (1) GCs.}
\begin{tabular}{lccc}
  \hline
  \hline
 & \multicolumn{2}{c}{Has GCs?} & Total \\
 & No (0) & Yes (1) & Galaxies \\ 
  \hline
Local Group Sample &  80 &  20 & 100 \\ 
  Georgiev Sample &   8 &  31 & 39\\ 
  Virgo Cluster Survey &   0 &  93 & 93 \\ 
  \hline
  \hline \\[-1.8ex] 
\end{tabular}

\label{tab:summarydata}
\end{table}

\begin{figure}
    \centering
    \includegraphics[width=0.49\textwidth]{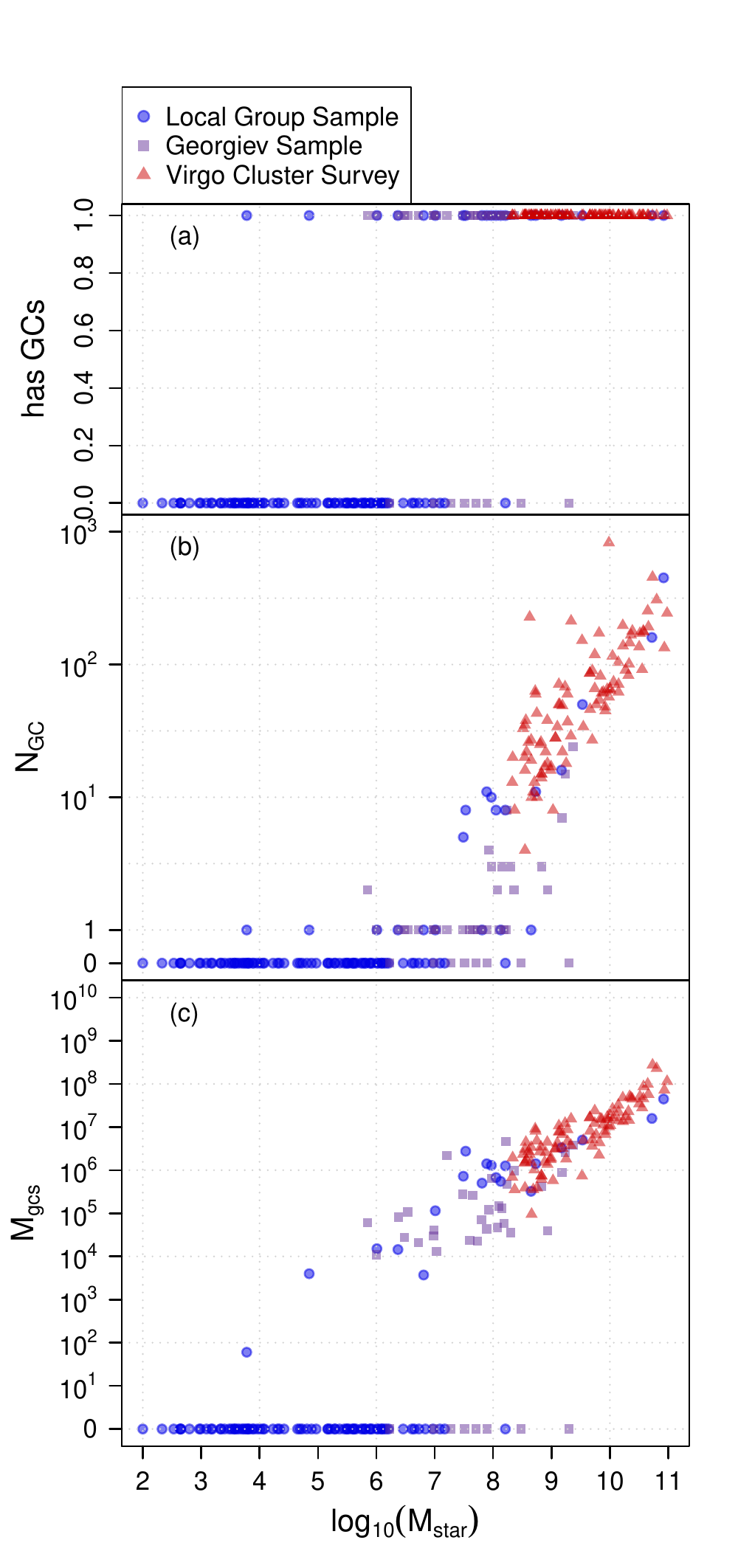}
    \caption{(a) The binary variable ``has GC'' ($y=0$ for no GCs, $y=1$ for has GCs) vs. $\log{M_{\star}}$, (b) the integer number of GCs $N_{GC}$ vs. stellar mass, and (c) the total GC system mass $M_{gcs}$ vs stellar mass.}
    \label{fig:dataonly}
\end{figure}

Figure~\ref{fig:dataonly} shows three of the response variables in our data set versus the total stellar mass $\log{M_{\star}}$: the three panels show (a) whether or not a galaxy has a GC, (b) the total number of GCs $N_{GC}$, and (c) the total mass in the GC system $M_{gcs}$. The colours and shapes of the symbols indicate the sample: the blue circles represent LG galaxies, purple squares represent the Georgiev (nearby dwarf) sample, and the red triangles represent the VCS galaxies.

It is immediately evident from Fig.~\ref{fig:dataonly}(a) that the transition from galaxies not having GCs to those that do is a gradual one, since there is extensive overlap of galaxies without ($y=0$) and with ($y=1)$ GCs in the mass range $M_{\star} \sim 10^6 - 10^7 M_{\odot}$.  In Figure~\ref{fig:dataonly}(b)
we observe that $N_{GC}$ generally increases with stellar mass, although $N_{GC}$ has notably high variance for galaxies between $10^8$ and $10^9M_{\odot}$. In Figure~\ref{fig:dataonly}(c), we observe a strong relationship between $M_{gcs}$ and stellar mass for the galaxies that have GCs, but we also note that galaxies without GCs span a large range of total stellar mass. 

In Figure~\ref{fig:dataonly}, the two lowest-luminosity galaxies with GCs are in the LG sample. These are Eridanus II at $M_{\star} = 7 \times 10^4 M_{\odot}$, holding a single cluster of $\sim 4000 M_{\odot}$,  and Ursa Major II at $M_* = 5.4 \times 10^3 M_{\odot}$, with a single cluster of just $\sim 60 M_{\odot}$. (See Appendix~\ref{app:photometry} for the relevant data on UMaII and new photometry of the cluster.) Contrarily, the highest-luminosity galaxy \emph{without} any GCs in the LG is the irregular IC1613 at $1.6 \times 10^8 M_{\odot}$  \citep[this small galaxy has a few star clusters, but no old GCs; see][]{wyder+2000}.

In Sections~\ref{sec:LR}-\ref{sec:hurdle}, we attempt to empirically model the relationships observed in Figure~\ref{fig:dataonly}, with the goals of quantifying and gaining insight into the properties of the observed transition region. In what follows, we always treat $\log{M_{\star}}$ as the predictor or covariate, and treat the other quantities (``has GC'', $N_{gc}$, $M_{gcs}$) as the response variables. In principle, we could use the halo (virial) mass $M_h$ of the galaxy as the predictor variable instead of $\log{M_{\star}}$, but the stellar mass has the strong advantage that it is more directly observable and also appears to have a simple relationship to $M_{gcs}$.

\section{Logistic Regression}\label{sec:LR}

To model the transition region using data for $M_{\star}$ and whether or not a galaxy has GCs (i.e., Figure~\ref{fig:dataonly}a), we can use logistic regression. Logistic regression provides an estimate of the \emph{probability} that a galaxy of given mass will contain a GC population. In Section~\ref{sec:logisticregression} we outline the basics of the model and define notation; a more complete introduction can be found in, for example, \cite{McCullaghNelder1989} and Chapter 5 of \cite{gelman2006data}.

Logistic regression is a specific kind of \textit{generalized linear model} (GLM). GLMs are an extensive and rich field of statistical models that have gone relatively underutilized in astronomy, but that hold much promise \citep[e.g., see the series of papers on GLMs starting with][]{deSouza2015overlookedI}.  Readers familiar with this particular GLM may prefer to skip ahead to Section~\ref{sec:firstresults}.

\subsection{Logistic Regression as a Generalized Linear Model}\label{sec:logisticregression}

Following the conventional statistics notation, we denote random variables with capital letters (e.g., $Y$) and observed quantities or realizations (i.e. observed data) of those random variables with lowercase letters (e.g. $y$). We adopt this convention here, defining $Y$ as a vector of the response or dependent random variable. Note that we also use capital $\mathbf{X}$ to define a matrix of \textit{predictor} variables or covariates, which are treated as known quantities (i.e. not random variables).

In a simple \textit{linear model}, we have
\begin{equation}
    E[\bm{Y}] = \mathbf{X}\bm{\beta}
    \label{eq:linear}
\end{equation}
where $\bm{Y}$ is a column vector of responses, $\bm{\beta}$ is a column vector of parameters,  $\mathbf{X}$ is a matrix with each row $i$ representing an observation and each column $j$ representing a particular predictor or covariate, and $E[]$ is the expected value. We write the expected or mean value of the response variables $\bm{Y}$ in eq.~\ref{eq:linear} as $\mathbf{X\beta} = \bm{\mu}$, where $\bm{\mu}$ is a vector. A statistical model for $\bm{Y}$ includes a random component. For example, it is common to assume that $Y_i$ follows a normal distribution with mean $\mu_i$, with the $Y_i$ independent given $\bm{\mu}$ and having common variance $\sigma^2$, written as: $Y_i \sim \mathcal{N}(\mu_i, \sigma^2)$. 

Eq.~\ref{eq:linear} implies that $E[\bm{Y}]$ is a vector of real numbers, but what if we want to exploit the linear model framework in cases where the $\mu_i$ are restricted? A $generalized$ linear model \citep[GLM,][]{McCullaghNelder1989} uses a \emph{link function} $g()$ that maps the whole real number line to the proper domain for $E[Y]$, that is:
\begin{equation}
    g(\bm{\mu}) = \mathbf{X}\bm{\beta}.
\end{equation}
We would like to use this expanded GLM framework to model a binary response variable $Y$. For each galaxy, we define a response variable $Y$ which takes the value  $y=0$ (galaxy does not have GCs) or $y=1$ (galaxy has GCs). If the $Y_i$ are Bernoulli distributed with probability of ``success'' $P(y_i=1) = p_i$, then $E[Y_i] = p_i$. In this case, we can generalize the linear model using a link function $g()$ that maps the whole real number line to the interval $[0,1]$. In this paper, we use the \textit{logit} function:
\begin{equation}
    \ln \left( \frac{\mathbf{p}}{1-\mathbf{p}} \right) = \mathbf{X} \bm{\beta},
    \label{eq:logit}
\end{equation}
where $\bm{p}/(1-\bm{p})$ is the \textit{odds}. 

In our example, we are interested in the probability $p$ that a galaxy has GCs, given some predictor variables $\mathbf{X}$ such as stellar mass, halo mass, morphological type, etc. We can express eq.~\ref{eq:logit} in terms of the probabilities $\bm{p}$ using the inverse logit function $logit^{-1}$:
\begin{equation}
    \bm{p} = \frac{1}{1+e^{-\mathbf{X}\bm{\beta}}}.
    \label{eq:invlogit}
\end{equation}
For example, if there are two predictor variables $X_1$ and $X_2$, then eq.~\ref{eq:invlogit} becomes
\begin{equation}
    \bm{p} = (1+e^{-(\beta_0 + X_1\beta_1 + X_2\beta_2)})^{-1}.
    \label{eq:twopredictors}
\end{equation}
Interaction terms for predictor variables such as $X_3 = X_1X_2$ can also be included:
\begin{equation}
    \bm{p} =  (1+e^{-(\beta_0 + X_1\beta_1 + X_2\beta_2 + X_1X_2\beta_3)})^{-1}.
    \label{eq:interact}
\end{equation}

The vector of parameters $\bm{\beta}$ is typically estimated using maximum likelihood, where the likelihood is given by
\begin{equation}
   \mathscr{L}(\bm{\beta};y) =  \prod^n_i p_i^{y_{i}} (1-p_i)^{1-y_i},
\end{equation}
and where each $p_i$ is given by equation~\ref{eq:invlogit}. Maximizing the likelihood function with respect to $\bm{\beta}$ requires an iterative, numerical procedure. Because the likelihood function for logistic regression is from the exponential family and is approximately quadratic in the area of the maximum, a reliable numerical method for the optimization is Iteratively Re-weighted Least Squares \citep[IRLS,][]{McCullaghNelder1989}. In Section~\ref{sec:firstresults}, we use the \textit{glm} function in the R Statistical Software Language to perform the logistic regression, which uses the aforementioned method \citep{RmanualStats}. 

\begin{figure}
    \centering
    \includegraphics[width=0.47\textwidth]{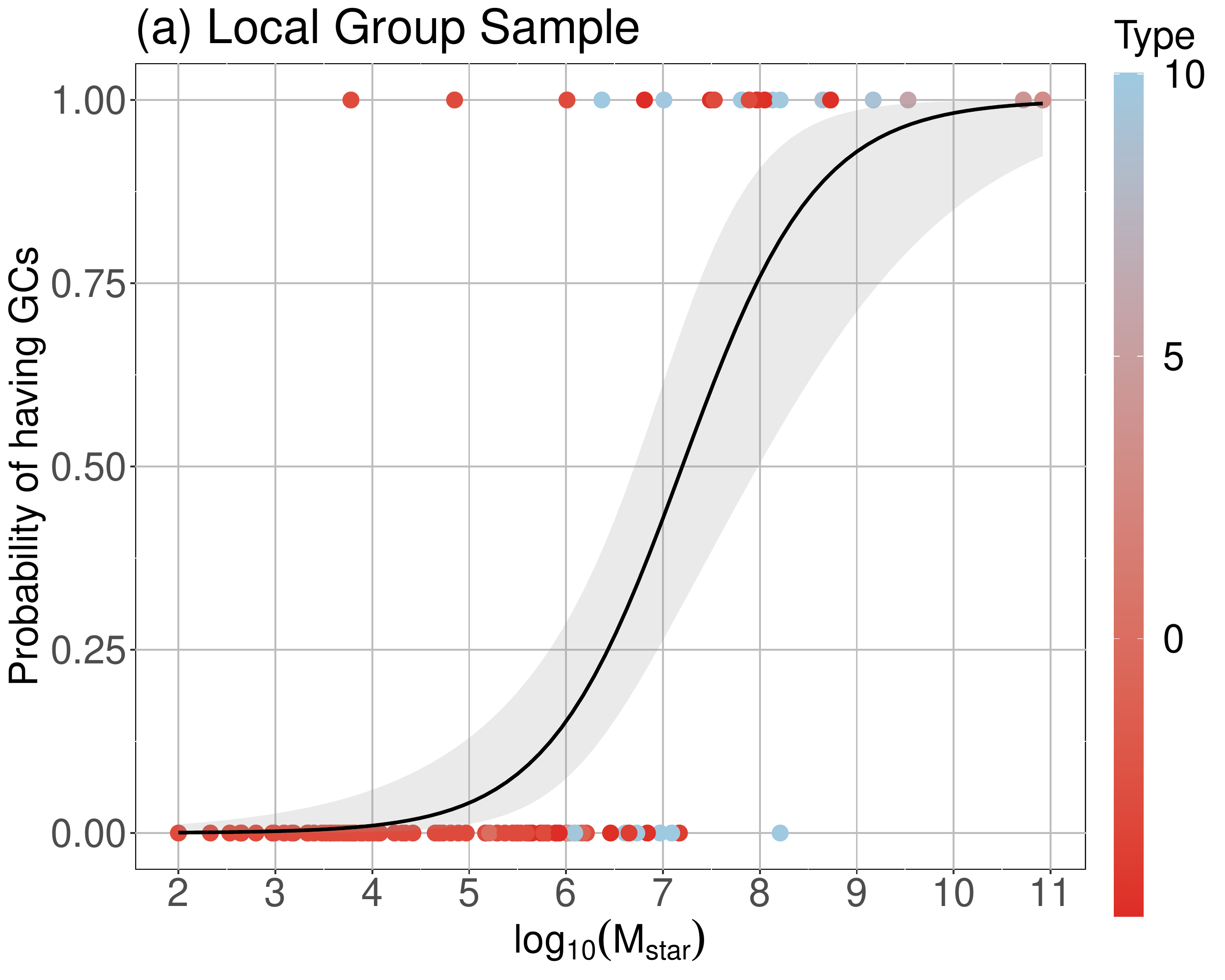}
    \includegraphics[width=0.47\textwidth]{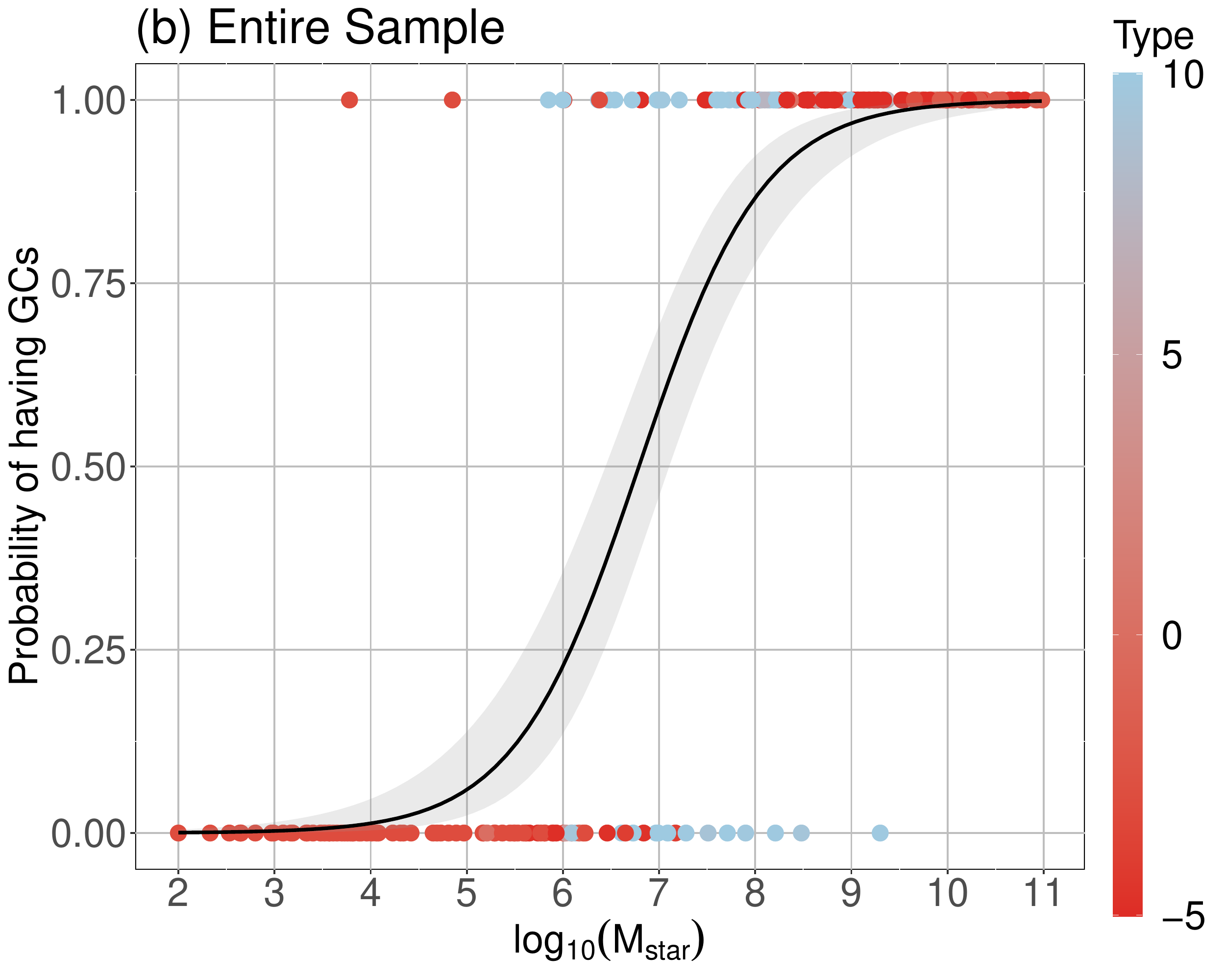}
    \caption{Logistic regression model assuming a single predictor (stellar mass) for (a) the Local Group, and (b) the entire sample. Galaxy morphological type from -5 (elliptical) to +10 (irregular) is color-coded as shown by the color bar, but is not included in the analysis (\textbf{see text}). The logit regression curve obtained through maximum likelihood is shown as the solid black line, and the grey regions show the estimated 95\% confidence intervals in probability for a given stellar mass.}
    \label{fig:logisticregression}
\end{figure}

\subsection{Application to the Local Group Sample}\label{sec:firstresults} 

Using logistic regression with a single predictor variable, the galaxy stellar mass $X = \log M_{\star}$, equation~\ref{eq:twopredictors} becomes
\begin{equation}
    \bm{p} = (1 + e^{-(\beta_0 + \beta_1 \log M_{\star})})^{-1},
    \label{eq:probGCs}
\end{equation}
where $\bm{p}$ represents the probability that a galaxy with stellar mass $\log M_{\star}$ has globular clusters. 

Using the LG sample only, the maximum likelihood estimate for $p$ as a function of $\log{M_{\star}}$ (eq.~\ref{eq:probGCs}) is shown in Figure~\ref{fig:logisticregression}(a) by the black sigmoid-shaped curve. The 95\% confidence interval is the grey region. The maximum-likelihood estimates for $\beta_0$ and $\beta_1$ are shown in the first column of Table~\ref{tab:logisticcoeff}. With this fit, the mass above which a galaxy has a greater than 50\% probability of having GCs is $M_{\star} = 10^{7.20} M_{\odot}$. But the transition region is broad:
$p= 0.1$ lies at $M_{\star} \approx 10^{5.67} M_{\odot}$, and $p=0.9$ at $10^{8.73} M_{\odot}$.

The corresponding \emph{halo mass} $M_h$ to our estimated point at which $p=0.5$ depends on which particular $M_{\star} \rightarrow M_h$ transformation is adopted.  For the frequently used double power-law form \citep[e.g.][]{hudson+2015,moster+2010}, $M_{\star}(p=0.5)$ corresponds to $M_h \simeq 1.1 \times 10^{10} M_{\odot}$.  If instead we adopt the three-part fitting function of \citet{behroozi+2013}, which is shallower
at the low-mass end, we obtain a slightly smaller value of $M_h \simeq 0.8 \times 10^{10} M_{\odot}$. In both cases, extrapolation of their relations is required, since the smallest
systems in their studies have $M_{\star} \sim 10^8 M_{\odot}$ or $M_h$ of a few $\times 10^{10} M_{\odot}$.

\subsection{Application to the Entire Sample}\label{sec:moredata}

The logistic regression for all galaxies in the Local Group, the Georgiev Sample (nearby dwarfs), and VCS, is shown in the bottom panel of Figure~\ref{fig:logisticregression}, with parameter estimates listed in the \textit{Entire Sample} column of  Table~\ref{tab:logisticcoeff}. With the entire sample, we can see that the mass at which a galaxy would have a 50\% probability of having GCs is now $\log{M_{\star}}=6.79$. This is notably similar to the result in \citet{georgiev+2010} of $\log{M_{\star}} \simeq 6.8 \pm 0.2$(Section~\ref{sec:dwarfs}). The logistic regression fit allows us to predict, for example, that a galaxy with $M_{\star} = 10^8 M_{\odot}$ would have a 76\% probability of having GCs.

\begin{figure}[!b]
	\centering
	\includegraphics[width=0.47\textwidth]{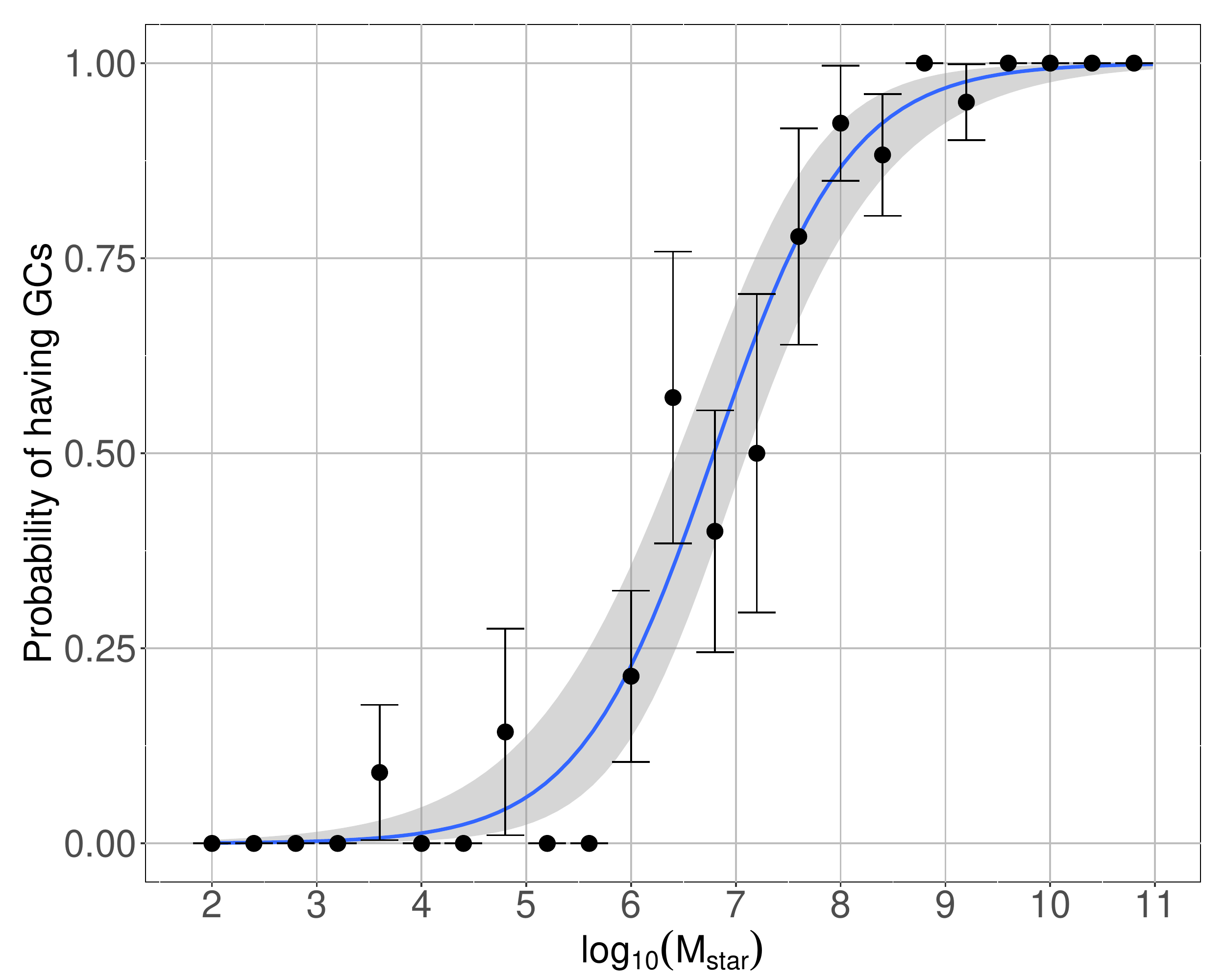}
	\caption{The data points show the observed fraction of galaxies that have GCs, plotted in bins of  $\Delta {\rm log} M_* = 0.4$.  The error bars represent 
		one standard deviation, estimated using method of moments and assuming a binomial distribution.
		The logistic regression fit from the bottom panel in Figure~\ref{fig:logisticregression} is superimposed
		to show the good correspondence between it and the data.  However, the curve was not derived as a fit to the binned data; see text.}
	\label{fig:logisticbin}
\end{figure}

Ideally the logistic regression curve should give a fair representation of the probability that an ``average'' galaxy of mass $\log{M_*}$ has GCs. A check of how well this model fits the data is shown in Figure \ref{fig:logisticbin}. Here, we display the fraction of galaxies that have GCs, plotted within bins of $\Delta {\rm log} M_* = 0.4$. The logistic regression curve and confidence intervals from the bottom panel of Figure~\ref{fig:logisticregression} are superimposed.  The curve represents fairly accurately what is happening through the transition region, taking into consideration the error bars and binning.
 
The summary of our findings using the logistic regression method is listed in Table~\ref{tab:logisticcoeff}, for the Local Group alone and then for the entire sample.
\begin{table}[!b] \centering 
	\caption{Logistic regression coefficients} 
	\label{tab:logisticcoeff} 
	\begin{tabular}{@{\extracolsep{5pt}}lcc} 
		\\[-6ex]\hline 
		\hline
		& \multicolumn{1}{c}{Local Group} & \multicolumn{1}{c}{Entire Sample} \\ 
		\hline \\[-1.8ex] 
		$\beta_0$ & -10.31 & -10.50 \\ 
		& (-14.59,-6.03) & (-13.33,-7.67) \\ 
		$\beta_1$ & 1.43 & 1.55 \\ 
		& (0.79,2.07) & (1.15,1.94) \\ 
		\hline \\[-3ex] 
		Observations & 100 & 232 \\ 
		\hline 
		\hline \\[-3ex] 
		\multicolumn{3}{l}{\textit{Notes: values in brackets are 95\% confidence intervals.}} \\
	\end{tabular} 
\end{table}

\subsection{Using Multiple Predictors}\label{sec:multiplepredictors}

Up to this point, we have only considered the galaxy stellar mass as a predictor of the probability that a galaxy has a GC population. However, it is possible that other characteristics of galaxies might also affect whether or not a galaxy has a GC population. Such characteristics might include the morphological type, ellipticity, integrated color, global star formation rate (SFR), or mean age of the stellar population,  among other parameters. For example, we might expect that the relative number of GCs could be lower if the SFR at the earliest times --- and thus the mean population age --- were lower.  Any number of quantities can be added as predictors in logistic regression, either individually (eq.~\ref{eq:twopredictors}) and/or through interaction terms (eq.~\ref{eq:interact}).

As a trial run for multiple predictors, we used the numerical morphological Hubble type T (which runs from -6 for extreme ETGs to +10 for irregulars).  Here the ultra-faint dwarfs were somewhat arbitrarily assigned $T = -3$ since they are structurally small, ellipsoidal, old, and gas-free.  The result, either with or without an interaction term, indicated that the $T$ made no significant change to the solution.  However, the Hubble type is a relatively crude parameter and was originally built for large galaxies rather than the small dwarfs that dominate the sample we are analyzing.  We conclude only that the first-order galaxy type does not obviously have an influence on the transition region, but additional work is needed to explore this direction more thoroughly.  For this paper, our discussion concentrates on $M_*$ as the single driving predictor.

\section{Poisson Regression}\label{sec:poisson}

Logistic regression is helpful in empirically describing the location and extent of the transition region. Going beyond the simple presence or absence of GCs in a galaxy, a logical next step is to model the \emph{number} of GCs ($N_{GC}$). Empirically, the average $N_{GC}$ clearly increases smoothly with galaxy stellar mass for galaxies with $M_{\star}>10^8 M_{\odot}$ as seen in Figure~\ref{fig:dataonly}b \citep[also see][]{peng+2008,georgiev+2010,harris+2013}. For progressively smaller galaxies, the chances of possessing GCs will decrease and thus the number of GCs will decrease as well.  

Other studies have modelled GC counts within a particular stellar mass range with a Poisson model \citep[e.g.][]{huang_koposov2020}. In the absence of physically-motivated models to explain both (1)  the randomness of $N_{GC}$ at each $\log{M_*}$, as well as (2) the observed relationship between $\log{M_*}$ and $N_{GC}$, a Poisson regression model is arguably the simplest approach. In \emph{Poisson regression}, the average number of GCs $\lambda$ varies as a function of one or more predictor variables, such as $\log{M_{\star}}$. In principle, one could relate $\lambda$ to more complicated functions of $\log{M_*}$ or even other galaxy parameters, but here we will retain the same $\bm{X}\beta$ as in the previous regression.

\subsection{Analysis}\label{sec:PoisAnalysis}

Like logistic regression, Poisson regression is an example of a generalized linear model (GLM). The link function $g()$ (see Section~\ref{sec:logisticregression}), in this case, is the natural logarithm: 
\begin{eqnarray}
 g(\bm{\lambda}) &=& \mathbf{X} \bm{\beta}, ~ \text{or} \\  
 \ln{\bm{\lambda}} &=& \mathbf{X} \bm{\beta}\label{eq:loglambda},
\end{eqnarray}
where $\bm{\lambda}$ is a vector of \textit{expected} values of the number of GCs \textbf{(i.e., $E[Y_i] = \lambda_i$; $y_i=N_{GC}$)}, $\bm{X}$ is the matrix of predictors used previously (in this case, the predictor is $\log{M_{\star}}$), and
$\bm{\beta}$ is a vector of parameters (\citealt{McCullaghNelder1989}; see also \citealt{deSouza2015overlooked} for other astronomy applications). In the context of this study, eq.~\ref{eq:loglambda} is
\begin{eqnarray}
    \ln{\lambda_i} &=& \beta_0 + \beta_1\log{M_{\star,i}} \\
    \ln{E[Y_i]} &=& \beta_0 + \beta_1\log{M_{\star,i}},
    \label{eq:poisline}
\end{eqnarray}
where $E[Y_i]$ is the expected value of the number of GCs for a galaxy with $\log{M_{\star,i}}$.

The Poisson regression is fit by maximizing a Poisson likelihood with respect to $\bm{\beta}$:
\begin{eqnarray}
   \mathscr{L}(\bm{\beta};y) &=& \prod^n_i \frac{\lambda_i^{y_i}e^{-\lambda_i}}{y_i!} \\
   &=& \prod^n_i \frac{(\mathbf{X_i} \bm{\beta})^{y_i}e^{-(\mathbf{X_i} \bm{\beta})}}{y_i!},
   \label{eq:poislikelihood}
\end{eqnarray}
which we accomplish through the \texttt{glm} function in \textbf{R}. The maximum likelihood estimates of $\beta_0$ and $\beta_1$ and their 95\% confidence intervals are $\hat{\beta}_0=-7.47~(-7.72,-7.21)$ and $\hat{\beta}_1=1.21~(1.18,1.24)$ (Table~\ref{tab:Poissoncoeff}).

\begin{table}[!htbp] \centering 
  \caption{Poisson regression coefficients} 
  \label{tab:Poissoncoeff} 
\begin{tabular}{@{\extracolsep{5pt}}lc} 
\\[-6ex]\hline 
\hline \\[-1.8ex] 
 & \multicolumn{1}{c}{\textit{Dependent variable: $N_{GC}$}} \\ 
\cline{2-2} 
 & Entire Sample \\ 
\hline \\[-1.8ex] 
 $\beta_0$ & -7.47 \\ 
  & (-7.72,-7.21) \\ 
 $\beta_1$ & 1.21 \\ 
  & (1.18,1.24) \\ 
\hline \\[-1.8ex] 
Observations & 232 \\ 
\hline 
\hline \\[-3ex] 
\multicolumn{2}{l}{\textit{Notes: values in brackets are 95\% confidence intervals.}} \\
\end{tabular} 
\end{table} 

\begin{figure}
    \centering
    \includegraphics[width=0.49\textwidth]{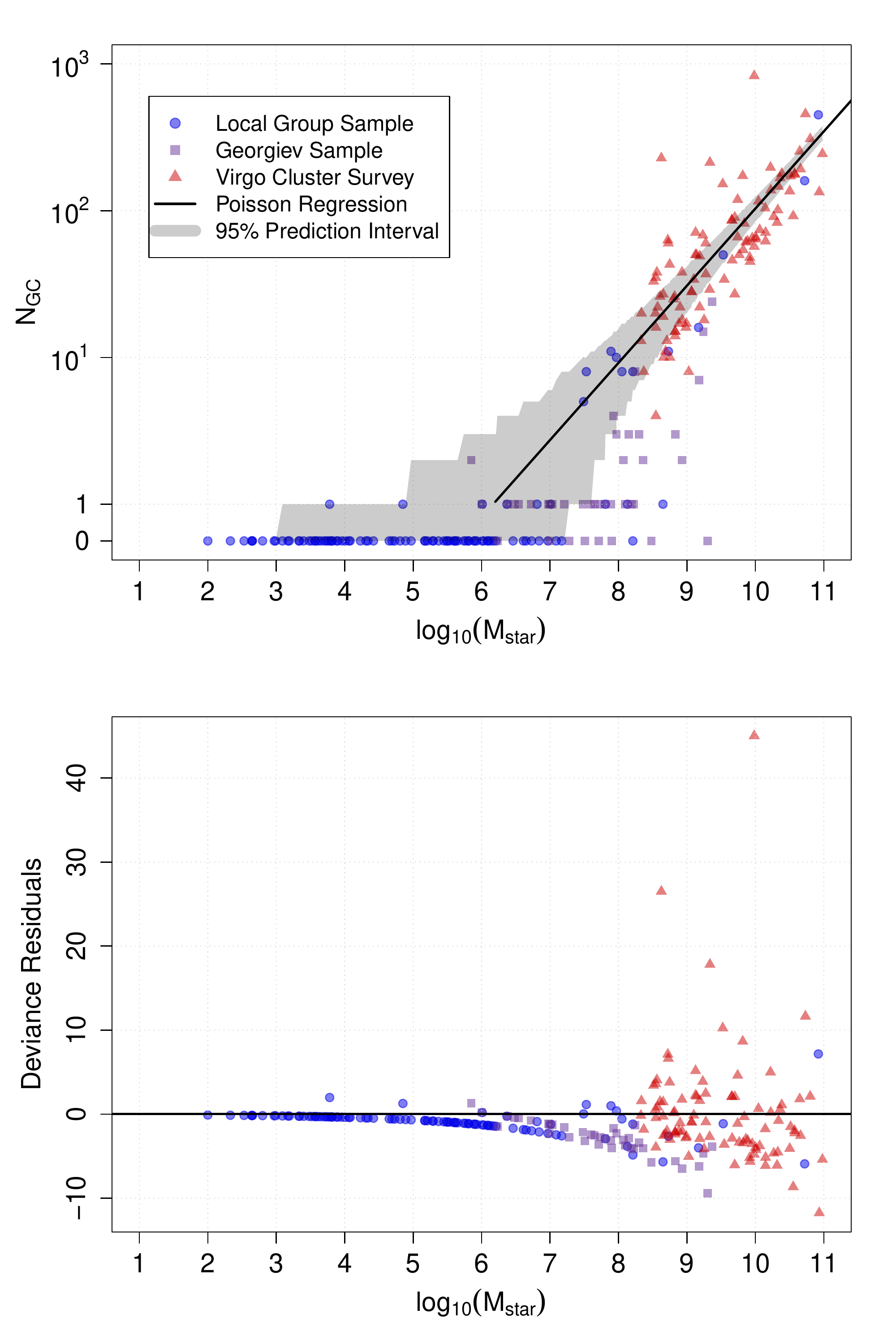}
    \caption{\textit{Top Panel:} The number of GCs in each galaxy as a function of galaxy stellar mass (points) and the Poisson regression fit (solid line). The line is not extended to $N_{GC}=0$ because of the log scale. The grey region signifies the \textit{prediction interval (not the confidence interval) from the Poisson model fit. The lack of real data with $N_{GC}=2$ or 3 is one indication the model is a poor description of the data.} \textit{Bottom panel:} Deviance residuals from the Poisson regression, as described in Section~\ref{sec:poisson}. The excess of negative deviance residuals is another indication of poor model fit (Section~\ref{sec:devianceresiduals}). Even when only data below $\log{M_{\star}}<8$ are used in the regression, the excess of negative deviance residuals persists. The dispersion test (Section~\ref{sec:dispersiontest}) implies the data set (and the data below $\log{M_{\star}}<8$) is overdispersed (see text).}
    \label{fig:poisson}
\end{figure}

The top panel in Figure~\ref{fig:poisson} shows the number of GCs as a function of the stellar mass for our data set, and the black curve shows the maximum likelihood estimate of the Poisson regression (i.e., using $\hat{\beta}_0$ and $\hat{\beta}_1$ in eq.~\ref{eq:poisline}). The black curve is not shown below $N_{GC}=1$ because of the logarithmic scale. The grey region in the top panel shows the 95\% \textit{predictive interval} (see Section~\ref{sec:predictive}) of the Poisson regression, while the bottom panel shows the \textit{deviance residuals} (see Section~\ref{sec:devianceresiduals}).

An important feature in the top panel of Figure~\ref{fig:poisson} is the lack of galaxies with 1, 2, 3, or 4 GCs in the transition region ($10^5-10^9 M_{\odot}$):  if such galaxies were present, some would lie on, above, and to the left of the regression line (and predominantly in the grey region; also see Section~\ref{sec:predictive}), whereas we see the majority of points are below and to the right. Even when the Poisson model is fit only to galaxies of $<10^8M_{\odot}$, this pattern in the residuals remains.  It is important to note that the position of the fitted line in Fig.~\ref{fig:poisson} is determined not just by the higher-mass galaxies but also by the large numbers of points at $N_{GC}=0$. 

This evidence suggests that the Poisson regression model in its simplest form is not a good description of the data, and especially not for the range below $<10^8M_{\odot}$. Said differently, the key anomaly is that not enough of the lowest-mass dwarfs have more than one GC. If at earlier epochs these dwarf galaxies did have multiple GCs, and \emph{if the counts followed a Poisson model at that time}, then the observations today are suggestive of physical evolution factors at work.  In Section \ref{sec:conclusions} below, we return to this point. 

In the next three subsections, further quantitative checks of the Poisson model are presented.

\subsection{Dispersion Test}\label{sec:dispersiontest}

The Poisson distribution has variance equal to the mean, and as stated above, we would expect this property to hold if the Poisson GLM is a good description of the data. It is common to test this assumption against the alternative that the variance is not equal to the mean -- this is known as a \textit{dispersion test}. Explicitly, we test the hypothesis
\begin{equation}
    \text{Var}[Y_i]=\lambda_i,
    \label{eq:pois}
\end{equation}
against the alternative that
\begin{equation}
    \text{Var}[Y_i] = \lambda_i + \alpha\lambda_i
    \label{eq:disppois}
\end{equation}
\citep{CAMERON1990347,cameron2013regression,cameron2005microeconometrics}, where $\alpha$ is a dispersion parameter to be estimated. If the data are \textit{equidispersed} then $\alpha=0$, consistent with the null hypothesis in equation~\ref{eq:pois}. If $\alpha>0$ then an \textit{overdispersed} Poisson model is favored (i.e., the variance is greater than the mean), whereas $\alpha<0$ implies  an underdispersed Poisson model. This test was done using the \texttt{AER} package and its \texttt{dispersiontest} function \citep{KleiberZeileis2008}.

The result of the dispersion test is in favour of the alternative: the estimate of the dispersion parameter for the entire sample of galaxies is $\alpha=43.4$ with a p-value of 0.04. When we apply the same model to only the galaxies with $M_{\star}\leq 10^8 M_{\odot}$, we find $\alpha=1.44$ with a p-value of 0.02, also significantly different from $\alpha=0$. The overdispersion suggested by this result is illustrated and explained in the following sections.

\subsection{Deviance Residuals}\label{sec:devianceresiduals}

One way to evaluate the appropriateness of a Poisson regression model (and other GLMs) is to look at the \textit{deviance residuals}. Although deviance residual plots are less commonly used in astronomy, they are quite common in other fields and are a useful tool for evaluating the model fit.

The deviance is a measure of \textit{lack-of-fit}\footnote{Put another way, the larger the deviance, the poorer the fit}, and the deviance residual is the square root of the contribution of a single data point to the deviance \citep{McCullaghNelder1989}:
\begin{equation}
    r_{D,i} = \text{sign}(y_i-\mu_i)\sqrt{\text{deviance}},
    \label{eq:rD}
\end{equation}
where $\mu_i$ is the expected value for point $i$. The mathematical form of the deviance is defined by the likelihood.

Examining the deviance residuals is a conventional way to assess a GLM. In the case of a Gaussian likelihood, the deviance residuals in eq.~\ref{eq:rD} would be $\text{sign}(y_i-\mu_i) \sqrt{(y_i-\mu_i)^2}=y_i-\mu_i$, so the deviance residuals and the ordinary residuals are the same.  In the case of a Poisson likelihood, the deviance residuals are $\text{sign}(y_i-\mu_i) \sqrt{2y_i\ln(y_i/\mu_i)-(y_i-\mu_i)}$ \citep{McCullaghNelder1989}. When the data match the expected value, the deviance residual is zero.

From examination of the deviance residual plot --- Figure \ref{fig:poisson}, bottom panel --- we notice two issues with the Poisson GLM. First, the residuals are much larger at high galaxy mass, which is suggestive of overdispersion in this range. Second, there is a long run of negative residuals between approximately $10^5$ and $10^8M_{\odot}$; within this range, there are fewer galaxies with small numbers of GCs than would be expected from a Poisson process.

\subsection{Predictive Checks \& Simulated Data}\label{sec:predictive}

\begin{figure}
    \centering
    \includegraphics[width=0.49\textwidth]{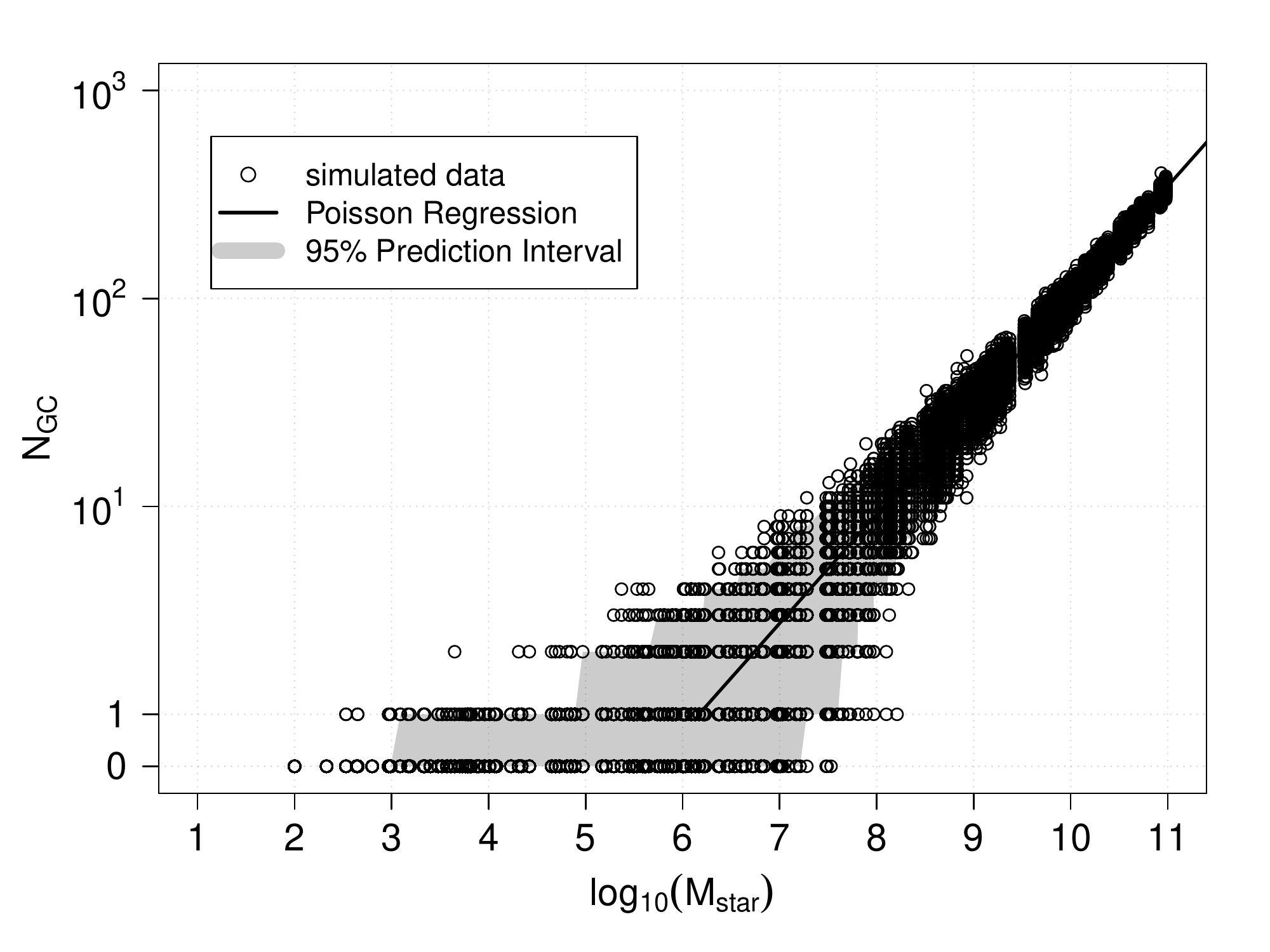}
    \caption{Simulated data of GC counts, $N_{GC}$, as a function of stellar mass, based on the Poisson model fit. The grey regions show the 95\% \textit{prediction interval} (not confidence interval) for the model, and the points are 100 independent realizations of simulated data sets of 232 galaxies each.}
    \label{fig:sims}
\end{figure}

Predictive checks and simulated data are also helpful in assessing model appropriateness. The 95\% prediction intervals for the Poisson regression are shown as grey regions in Figures~\ref{fig:poisson} and \ref{fig:sims}. These intervals represent the region in which 95\% of the data should be found if the model were correct. As we can see in the top panel of Figure~\ref{fig:poisson}, most of the real galaxies between $10^6$ and $10^9M_{\odot}$ are found outside and to the right of this region.

To strengthen our argument, we also simulate 100 independent data sets of 232 galaxies each using our Poisson regression fit and the galaxy stellar masses of our real galaxies. These simulated $N_{GC}$ values given $M_{\star}$ are shown in Figure~\ref{fig:sims}, along with the prediction intervals and the regression line from the fit to the real data. Again, we see how the data \textit{should} be distributed if they were to follow this Poisson model:  the distribution of simulated points in Figure~\ref{fig:sims} is quite different than the distribution of real data in the top panel of Figure~\ref{fig:poisson}.

\begin{figure}
    \centering
    \includegraphics[width=0.38\textwidth]{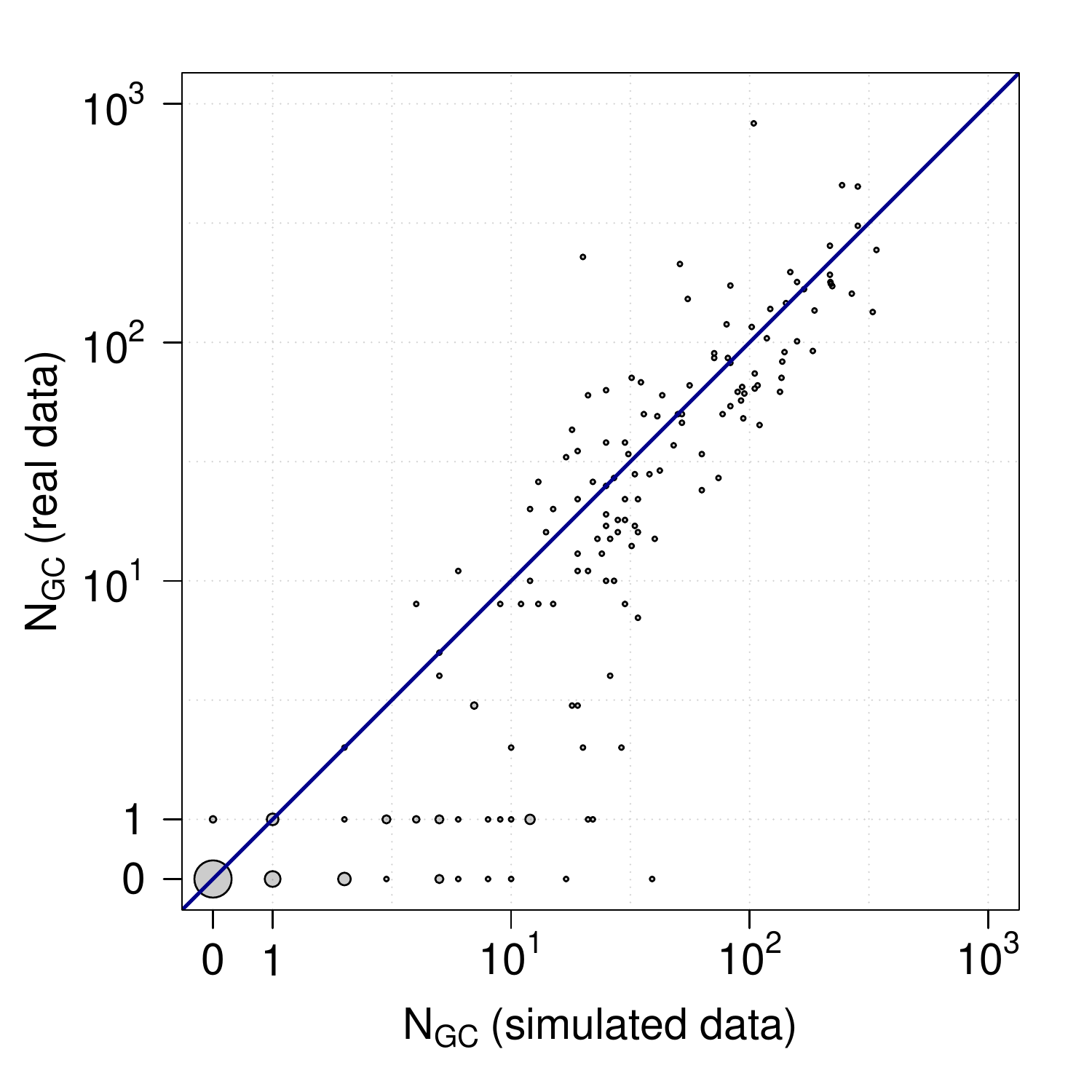}
    \caption{Comparison between the simulated $N_{GC}$ values from the Poisson regression model to the real data values of $N_{GC}$. Point locations correspond to the number of GCs in the simulated data (horizontal axis) and the number of GCs in the real data (vertical axis) for a given galaxy mass. Circle areas are proportional to the number of pairs of simulated and real $N_{GC}$ values at that location. For example, there are many pairs of $N_{GC}$ (simulated, real) at (0,0). If the Poisson model were a good description of the data, then the points should be scattered about the one-to-one line shown in dark blue. However, a lack of real data points with $N_{GC}=1,2,3,\text{and }4$ is apparent in the lower left-hand region of the plot.}
    \label{fig:realVSsiM}
\end{figure}

Figure~\ref{fig:realVSsiM} gives another perspective;  using one of our simulations, this figure compares the simulated data of GC counts for each galaxy to the actual data of GC counts for each galaxy. Each point  represents a pair of simulated $N_{GC, simulated}$ and real GC $N_{GC, real}$ data counts for a given galaxy. The size of the points are proportional to the number of pairs with the same $(N_{GC, simulated}, N_{GC, real})$. If the simulated data and real data were similar, then the points would follow the blue one-to-one line. However, there is a paucity of simulated and real data pairs above the blue line between $1-4$ GCs, and an excess of points below. This is another reflection of the conclusion stated above: the Poisson model cannot describe the data well. Too few of the small dwarfs in the real data have GCs at the present epoch.

\section{Linear Regression}\label{sec:mass}

Using only the numbers of GCs in a galaxy ignores the systematic changes in GC properties with host galaxy mass: both the mean GC mass and the standard deviation of the lognormal mass distribution increase systematically with galaxy luminosity \citep[e.g.,][]{villegas+2010,harris+2014}. Therefore, it is worth investigating how GC system mass ${M_{gcs}}$ relates to host galaxy stellar mass. 

In this section we explore a linear regression approach to model the relationship seen in Figure~\ref{fig:dataonly}c using the simple form
\begin{equation}
    \log M_{gcs} = \gamma_0 +\gamma_1 \log M_{\star},
\end{equation}
which is fit using unweighted least-squares to estimate $\gamma_0$ and $\gamma_1$. When performing this regression, we first exclude galaxies without GCs.  But in the next section, we add back the ones with no GCs and use a \textit{hurdle model} to empirically describe the $M_{gcs} - M_{\star}$ relation.

We perform the regression first on the LG Sample, and then on the entire sample; Table~\ref{tab:linearcoeff} summarizes the coefficient estimates and their 95\% confidence intervals. The estimated slope $\gamma_1$ is slightly but not appreciably different across these two cases when considering the 95\% confidence intervals.

\begin{figure*}
    \centering
    \includegraphics[width=.8\textwidth]{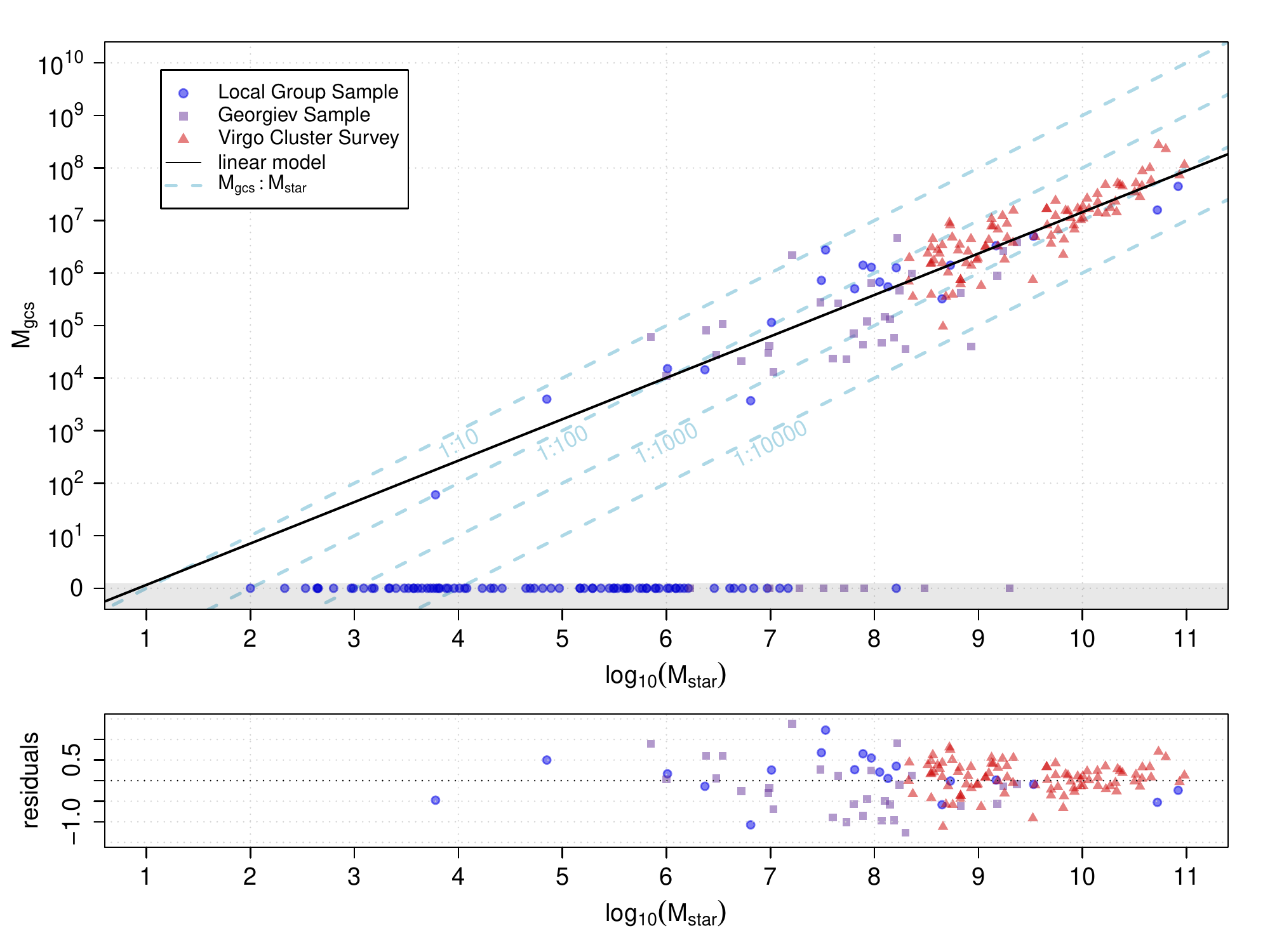}
    \caption{\emph{Upper panel:} Empirical relationship between the mass in GCs and the total stellar mass of the host galaxy. Members of the Local Group are blue circles, Virgo Cluster Survey members are red triangles, and other nearby dwarfs (see text) are purple squares.  The points along the bottom are the galaxies with no GCs. The solid line shows an unweighted linear regression of $\log_{10}(M_{gcs}/M_{\odot})$ on $\log_{10}(M_{\star}/M_{\odot})$ for the entire sample. The dashed light blue lines correspond to the lines along which 10\%, 1\%, 0.1\%, and 0.01\% of the galaxy's stellar mass would be made up of GCs (see text).
    \emph{Lower panel:} Residuals from the linear regression.}
    \label{fig:mass}
\end{figure*}

Figure \ref{fig:mass} shows $M_{gcs}$ versus $M_{\star}$ for the entire sample and the linear regression fit (solid black line). The excluded galaxies are indicated by the grey regions. In the bottom panel of this plot are the residuals, which have a slightly larger variance for galaxies of lower stellar mass.

\begin{table}[!htbp] \centering 
  \caption{Linear regression coefficients} 
  \label{tab:linearcoeff} 
\begin{tabular}{@{\extracolsep{5pt}}lcc} 
\\[-3ex]\hline 
\hline \\[-1.8ex] 
 & \multicolumn{2}{c}{\textit{Dependent variable: $\log M_{gcs}$ }} \\ 
 \hline
 & Local Group & Entire Sample \\ 
\hline \\[-1.8ex] 
 $\gamma_0$ & $-$0.381 & $-$0.725 \\ 
  & ($-$1.487,0.725) & ($-$1.283,$-$0.166) \\ 
  & & \\ 
 $\gamma_1$ & 0.756 & 0.788 \\ 
  & (0.617,0.895) & (0.726,0.851) \\ 
  & & \\ 
\hline \\[-1.8ex] 
Galaxies & 19 & 143 \\ 
\hline 
\hline \\[-3ex] 
\multicolumn{3}{l}{\textit{Note: Values in brackets are 95\% conf. intervals}} \\ 
\end{tabular} 
\end{table}

\begin{figure}
    \centering
    \includegraphics[width=0.49\textwidth]{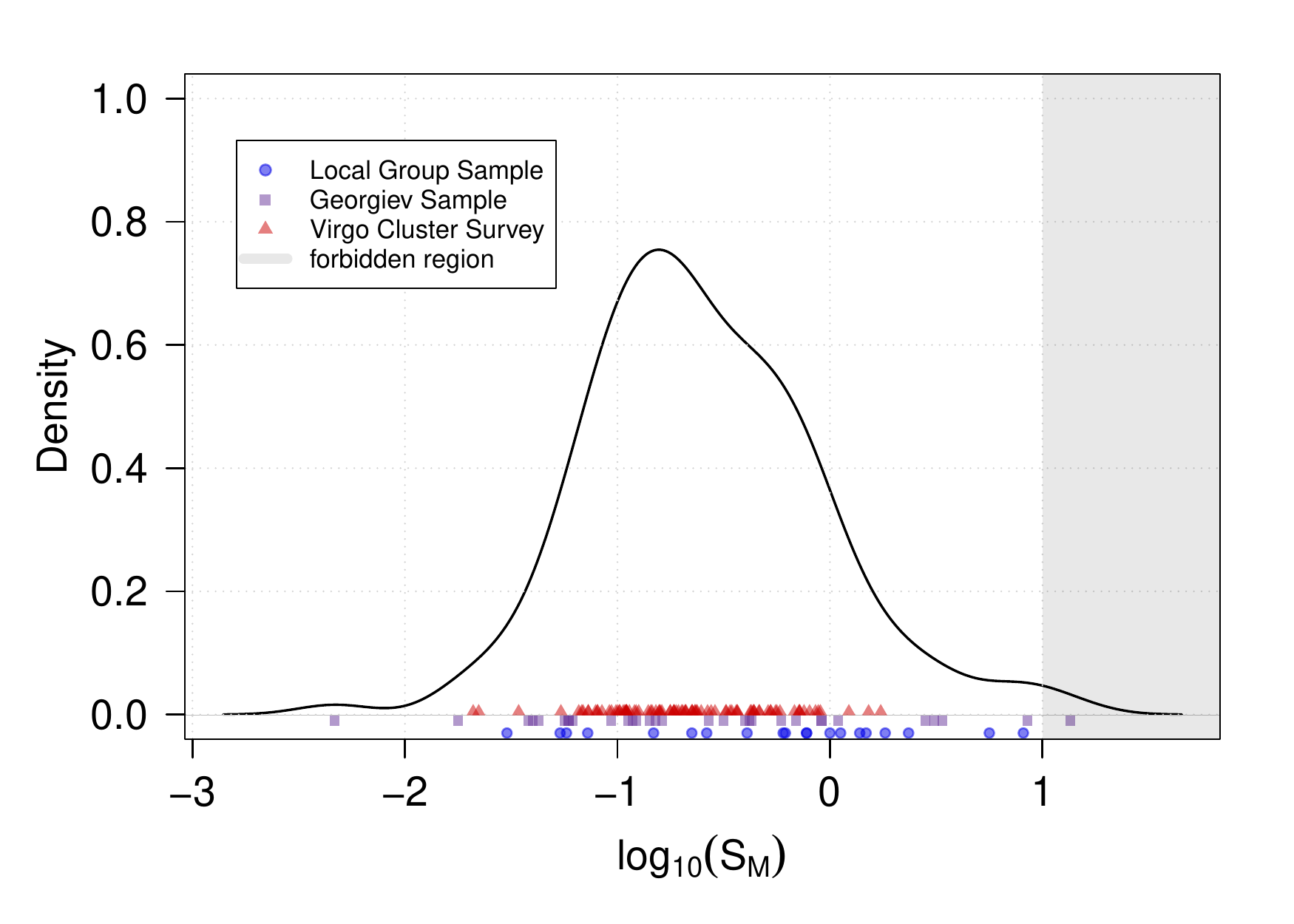}
    \caption{Kernel-smoothed distribution of the mass ratio $S_M=100*(M_{gcs}/M_{\star})$. The shaded area at right indicates the approximate ``forbidden zone'' of $log(S_M)\gtrsim1$ expected from recent cluster formation theory; see text.}
    \label{fig:density}
\end{figure}

The relationship between $M_{gcs}$ and $M_{\star}$ is often represented as the \emph{specific mass} ratio $S_M = 100(M_{gcs}/M_{\star})$ \citep{peng+2008}. The fact that $M_{gcs}$ does not vary in direct proportion to $M_{\star}$ guarantees that $S_M$ will not be constant with $M_{\star}$:  from the results above, we expect $S_M \sim M_{\star}^{-0.25}$ for galaxies that have GCs.

Extrapolating the mean relation downward, a physically interesting point is reached at $M_{\star} \simeq 10^4 M_{\odot}$ where $S_M = 10$, i.e.~a level where 10\% of the host galaxy stellar mass would be taken up by its GCs (if it had any). According to recent star cluster formation simulations \citep{howard+2018,lahen+2020,ma+2020,chen_vogelsberger2021}, a newly formed star cluster is unlikely to take up more than $\sim$10\% of its host Giant Molecular Cloud (GMC) mass, which would correspond to an approximate maximum $S_{M}$ of $\sim 10$ within broad limits. The lower edge of this $S_M > 10$ zone is marked by the blue dashed line labeled 1:10 in Figure~\ref{fig:mass} and by the shaded region in the marginal distribution of $\log(M_{gcs}/{M_{\star}})$ in Figure \ref{fig:density}. Interestingly, the $S_M$ values for the real galaxies in our sample fall below this limit. The five highest-$S_M$ cases observed are Eridanus II ($S_M=5.6$), VCC1363 (6.2), Sagittarius (8.1), KK27 (8.5), and UGC8638 (14). Other recent work for UDGs (Ultra-Diffuse Galaxies) in Virgo
and Coma, however, indicates that particularly cluster-rich examples of this class of galaxies may exceed $S_M = 10$, though the estimates are admittedly uncertain \citep{forbes+2020}.

There are two additional evolutionary effects that could influence $S_M$. First, the smallest galaxies may have had much of their initial gas content expelled by Type II SNe at an early stage, thus preventing later star formation and increasing $S_M$ if the GCs formed earliest in the star formation burst \citep[e.g.][]{mclaughlin1999}. Second, the long-term effects of dynamical erosion on the cluster mass (particularly stellar evaporation, disk shocking, dynamical friction, and tidal dissolution) will have the opposite effect of lowering $S_M$ with time; cf. the discussion in Section~\ref{sec:conclusions}. We can speculate that objects like UMaII and EriII are so small that they may have originated from single original seed halos (at a minuscule $M_h \sim 10^{7-8} M_{\odot}$) with no further important mergers.

The relationship between $S_M$ and $M_*$ has been extensively examined \citep{mclaughlin1999,peng+2008,georgiev+2010}.  \citet{mclaughlin1999} proposed a ``universal'' formation efficiency of $S_M = 0.26 \pm 0.05$ based on galaxies with $M_* \gtrsim 10^{10} M_{\odot}$ (at the high end of Figure~\ref{fig:mass}). \citet{peng+2008} and \citet{georgiev+2010} obtained similar estimates for $M_* \gtrsim 2 \times 10^8 M_{\odot}$. Although some of these analyses include gas mass in the mass ratio and others do not, the resulting estimates are similar except for the very biggest ETGs with huge X-ray halos (such cases are not included here). An $S_M$ level of $\sim 0.2$ would sit between the two middle blue dashed lines in Figure~\ref{fig:mass} that bracket most of the observed galaxies for $M_* \gtrsim 10^8 M_{\odot}$. 

While the linear regression provides some insight, its major drawback is that galaxies without GCs are excluded from the analysis, leaving out the additional binary information about whether or not a galaxy has a GC population. This motivates the next section, where we introduce an empirical \textit{hurdle model} that includes all galaxies, their stellar masses, and the presence or absence of GCs.

\vspace{-3ex}
\section{Bayesian lognormal hurdle model}\label{sec:hurdle}

As noted above, modelling the trend of $\log{M_{gcs}}$ versus $\log{M_{\star}}$ with a simple linear regression ignores the galaxies without GC populations, but a logistic regression model ignores the relationship between ${M_{gcs}}$ and $M_{\star}$ for galaxies that do have GC populations. We would like to model galaxies with and without GCs, and include the $M_{gcs}$ information, simultaneously in a single statistical model.

Many fields using applied statistics, such as economics, biostatistics, and environmental science, make use of \textit{hurdle models} and \textit{zero-inflated models} to model data that contain significant numbers of zeros (in our case, these would be the many low-luminosity galaxies with no GCs). Hurdle models describe processes for which the value of the response variable is zero until a critical or transition value of the covariate is reached. After this ``hurdle'' is overcome, the data then follow some other distribution. Zero-inflated models describe situations in which there are many more zeros in the distribution of the response variable than can be described by standard distributional assumptions. We provide a short list of hurdle and zero-inflated model applications in other fields to demonstrate their wide usage \citep[e.g., biostatistics studies about vaccines and HIV, ecological and environmetrics studies about Harbor seals and soil pollution, insurance studies in actuarial science, and transportation studies about traffic crashes, ][]{rose2006use,hu2011zero,ver2007space,2019EnST...53.6824H,boucher2007risk,ma2015modeling}. 

For the problem at hand --- where we are interested in the transition mass region of galaxies hosting or not hosting GCs --- a hurdle model seems appropriate.  Hurdle models are another example of GLMs, which more broadly have promise for astrophysical modelling \citep[see for example,][]{deSouza2015overlookedI, elliott2015overlooked,deSouza2015overlooked,Hattab2019MNRAS,Lenz2016A&A...586A.121L,hilbe2017bayesian}.

We implement a \textit{lognormal hurdle model} using a Bayesian approach. In this model, the GC system mass $M_{gcs}$ follows a lognormal distribution if the GC system exists, and the probability of existence is governed by the logistic model described in Section~\ref{sec:logisticregression}. Under this model, the \textit{expected} or mean value of the mass in GCs is given by
\begin{equation}
    E[\log{M_{gcs}}] = \left( \frac{1}{1 + e^{-(\beta_0 + \beta_1 \log {M_{\star}})}} \right)(\gamma_0 + \gamma_1\log {M_{\star}}).
    \label{eq:BayesHurdle}
\end{equation}
The first term in eq.~\ref{eq:BayesHurdle} is the inverse logistic function (i.e., eq.~\ref{eq:invlogit}) while the second term is the linear portion of the model. Together, these provide a gradual transition from no GCs to having GCs, which is referred to as the `hurdle'. The four free parameters
$(\beta_0,\beta_1)$ and $(\gamma_0,\gamma_1)$ are the coefficients for the logistic and linear portions of the model respectively.  The structure of this solution resembles a similar Bayesian lognormal hurdle model used
by \citet{hilbe2017bayesian} (see their Chapter 7), where they model the correlation between
galaxy stellar mass and halo mass predicted from simulations.

We use the \textit{Bayesian Regression Models using Stan} (\texttt{brms}) package \citep{brmsBurkner, BurknerRJournal} in the R Statistical Software Environment \citep{RSoftware} to implement the hurdle model for this study. The Bayesian lognormal hurdle model is built into this package through the general function \texttt{brm}, making it easy to implement in essentially one line of code.\footnote{On the first author's 4-year old laptop, we ran four sequential, independent Markov chains using the \textit{brms} package; the \textit{Stan} compiler and model run took $\sim49$ seconds. We show example code in Appendix~\ref{app:hurdle}.} 

The default priors in \texttt{brm} are flat, improper priors on the model parameters, although these can be changed by the user to different (proper) prior distributions. We opted to try both the default and our own diffuse priors. For the latter, we used the following:
\begin{align}
     \beta_0 &\sim \mathcal{N}(-13,5) \label{eq:beta0prior} \\
     \beta_1 &\sim \mathcal{N}(1, 5) \label{eq:beta1prior} \\
     \gamma_0 &\sim \mathcal{N}(0,2) \label{eq:gamma0prior} \\
     \gamma_1 &\sim \mathcal{N}(1,2), \label{eq:gamma1prior}
 \end{align}
where the first argument is the mean, and the second is the standard deviation. The results using the priors above were essentially the same as for the improper, flat default priors. Thus, we report our results using eq.~\ref{eq:beta0prior}-\ref{eq:gamma1prior}, because in principle we prefer even weakly informative priors to improper priors when some prior information is available. For the lognormal part of the model, the parameter $\sigma$ in the lognormal distribution is assumed to be Student $t$-distributed with three degrees of freedom (the default in \texttt{brm}).
\begin{figure*}[!t]
	\centering
	\includegraphics[width=0.7\textwidth]{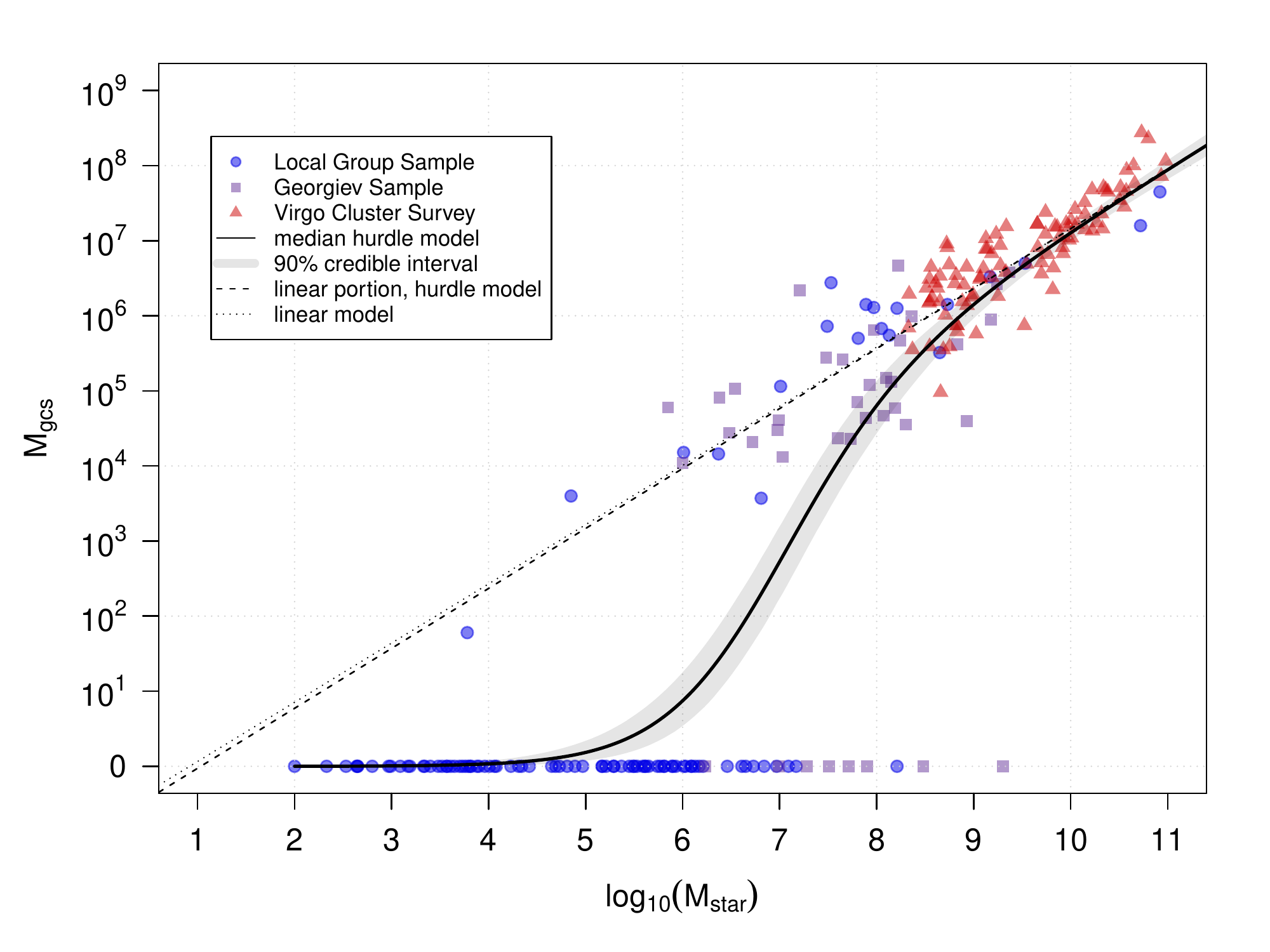}
	\caption{Bayesian lognormal hurdle model for the entire sample of galaxies. The solid black curve shows the mean estimate, and the grey regions show the 90\% credible interval. The linear portion of the hurdle model fit is shown with a dashed line, and for comparison the simple linear model is shown with a dotted line. (The vertical axis is shown in regular space in order to show cases when there are no GCS. As before, masses are in $M_{\odot}$ units.)  The linear model and the linear portion of the hurdle model are virtually identical, but the hurdle model has the advantage that it better describes the data within and below the transition region.}
	\label{fig:hurdle2}
\end{figure*}

It should be noted that the measurement uncertainties in $\log{M_{\star}}$, $N_{gc}$, and $\log{M_{gcs}}$ are not included in this model. Including these is non-trivial since the measurement uncertainties are heteroskedastic and non-Gaussian, so we leave this to future work.

Parameter summaries for equation~\ref{eq:BayesHurdle} are provided in Tables~\ref{tab:hurdlecoeffLG} and \ref{tab:hurdlecoeff} for the LG and entire sample respectively. The inferred value of $\beta_1$, which determines the sharpness in rise of the logistic regression, is similar for both the LG  and entire sample, although the 95\% credible interval is narrower in the latter. Introducing the Virgo and dwarf galaxies into the sample decreased the estimate of $\gamma_1$, but only slightly: the 95\% Bayesian credible intervals show considerable overlap (see the last row in Tables~\ref{tab:hurdlecoeffLG} and \ref{tab:hurdlecoeff}). 

Figure~\ref{fig:hurdle2} shows the hurdle model mean estimate (solid black curve) and 90\% Bayesian credible region (grey regions) given the entire sample of galaxies (points) and the prior parameter distributions in eqs.~\ref{eq:beta0prior}-\ref{eq:gamma1prior}. The expected value for $M_{gcs}$ in the transition region from $\sim 10^5$ and $10^9 M_{\odot}$ reflects that the populations of galaxies that have and don't have GCs overlap in this mass range.

The hurdle model results can be interpreted and used in a couple of ways. On the one hand, if we observe a galaxy and see that it has a GC population, then we can use the hurdle model fit to predict $M_{gcs}$ using the linear portion of the model.  For example, if we observed a galaxy in the transition region \emph{with a GC population}, and having mass $M_{\star}=10^7M_{\odot}$, we would predict its GC population mass to be $\simeq 10^{4.80}M_{\odot}$. On the other hand, if we consider a galaxy with $M_{\star}=10^7M_{\odot}$ but we do not know if it has a GC population, then the hurdle model expects the mass of the GC population to be $\simeq 10^{2.73}M_{\odot}$; this is because the probability of there being any GCs at all is only $\simeq 0.57$. 

\begin{table}
	\centering 
	\caption{Hurdle model coefficients for the LG sample} 
	\label{tab:hurdlecoeffLG} 
	\begin{tabular}{@{\extracolsep{5pt}} lccc} 
		\\[-3ex]\hline 
		\hline \\[-1.8ex] 
		& Estimate & Est.Error & 95\% Cred.Int. \\ 
		\hline \\[-1.8ex] 
		$\beta_0$ & $-11.53$ & $2.46$ & $(-16.97,-7.34)$ \\ 
		$\beta_1$ &  $1.60$ & $0.36$ & $(0.99,2.41)$  \\ 
		$\gamma_0$ & $-0.98$ & $0.66$ & $(-2.41,0.19)$  \\ 
		$\gamma_1$ & $0.82$ & $0.08$ & $(0.67,1.00)$ \\ 
		\hline \\[-1.8ex] 
		\multicolumn{4}{l}{\textit{Values in brackets are 95\% credible intervals}} \\ 
	\end{tabular} 
\end{table}

\begin{table}[!h]
	\centering 
	\caption{Hurdle model coefficients for entire sample} 
	\label{tab:hurdlecoeff} 
	\begin{tabular}{@{\extracolsep{5pt}} lccc} 
		\\[-3ex]\hline 
		\hline \\[-1.8ex] 
		& Estimate & Est.Error & 95\% Cred.Int \\ 
		\hline \\[-1.8ex] 
		$\beta_0$ & $-10.83$ & $1.50$ & $(-14.06,-8.23)$  \\ 
		$\beta_1$ & $1.59$ & $0.21$ & $(1.22,2.05)$ \\ 
		$\gamma_0$ & $-0.83$ & $0.29$ & $(-1.41,-0.29)$ \\ 
		$\gamma_1$ & $0.80$ & $0.03$ & $(0.74,0.86)$ \\ 
		\hline \\[-1.8ex] 
		\multicolumn{4}{l}{\textit{Values in brackets are 95\% credible intervals}} \\ 
	\end{tabular} 
\end{table} 

\section{Discussion and Comparisons}\label{sec:conclusions}

In this study, we have explored how the existence and mass of GC populations is related to galaxy stellar mass. Figs.~\ref{fig:logisticregression}, \ref{fig:mass}, and \ref{fig:hurdle2} are fundamentally similar: all show where the contribution of GCs to the galaxy stellar mass ramps downward and eventually vanishes.  In all versions, the conclusion is that for $M_{\star} \lesssim 10^6 M_{\odot}$, GCs are very unlikely to be present. The equivalent dark-matter
halo mass at that point \citep[from][]{behroozi+2013} is $M_h \sim 10^9 M_{\odot}$. While halo masses below that level may not be capable of producing sufficiently dense, massive star clusters that can survive to the present day, finding a \textit{massive} dark matter halo with a GC population but very little $M_{\star}$ would be very interesting.

Comparisons of this work can be made with the recent studies of \cite{forbes+2018}, who plot $M_{gcs}$ versus galaxy \emph{halo} mass $M_h$; and 
\cite{huang_koposov2020}, who plot GC \emph{numbers} versus galaxy luminosity (closely equivalent to stellar mass).  Forbes et al.~estimate $M_h$ for a sample of dwarf galaxies including the Local Group members.  For pressure-supported galaxies they use the velocity dispersion within one effective radius ($R_e$), correct for adiabatic contraction of the halo, and then extrapolate to the virial mass $M_{200}$ with an NFW-type halo profile model. For rotation-supported systems they use the rotation curve or line-width profile, again combined with an assumed NFW DM profile, to extrapolate to $M_{200}$.  These prescriptions require large radial extrapolations from the inner, directly observed stellar light. Their predicted virial masses for a given $M_{\star}$ for the smallest galaxies are roughly an order of magnitude lower than for other methods based on abundance matching or weak lensing \citep{moster+2010,behroozi+2013,hudson+2015,read+2017}. Forbes et al.~conclude (see their Figure 3) that the 10\% probability for containing a GC is near $M_h \lesssim 10^{9} M_{\odot}$ (by their calibration), which
would roughly correspond to $M_{\star} = 10^{5-6} M_{\odot}$. Given the different SHMRs used, their result is approximately consistent with ours: we find the probability of such galaxies containing a GC to be between 5 and 22\%.

\cite{huang_koposov2020} used the Gaia DR2 database to carry out a uniform search for GCs around the 55 Milky Way satellite dwarfs that are within 450 kpc.  They find no previously unknown ones down to very low GC luminosity limits (where the limit depends on the galaxy distance) and specifically account for GC detection incompleteness as a function of magnitude and distance. They conclude that there is a $>90$\%
probability of no GC for galaxies with $M_V > -9$, which would be equivalent to log $M_{\star} < 5.73$ for the same $(M/L_V)$ ratio we have used above. By comparison, we obtained $p=0.1$ at log $M_{\star} = 5.4$ ($M_h \simeq 10^9 M_{\odot}$), slightly less massive than their estimate, but not drastically so given the different database used in their study and their different statistical approach with binning.

As discussed in Section \ref{sec:poisson}, Poisson regression for $N_{GC}$ versus stellar mass does not match the data:  too many of the LG dwarfs in the range $10^5-10^8 M_{\odot}$ have no clusters, a discrepancy
that cannot be accounted for just with randomness in a simple Poisson model. Going beyond statistical effects, we should therefore not ignore the presence of astrophysical effects on GC formation and GC survival in the smallest galaxies.

What might drive the paucity of GCs in present-day galaxies below $M_{\star} \sim 10^7 M_{\odot}$? Both GC formation and subsequent dynamical evolution must be important factors. Observationally, the mean GC mass for galaxies \emph{at the present day} ($z=0$) is typically $10^5 M_{\odot}$ for large galaxies, but decreases systematically for the smallest dwarfs with considerable variance \citep[see][as well as the data presented here]{villegas+2010}. Recent theory that tracks GC formation and evolution within simulated galaxies \citep{reina-campos_kruijssen2017,pfeffer+2018} indicates that at time of formation in low-metallicity dwarfs (redshifts $z \gtrsim 4$), many clusters form in the range $10^{5-6} M_{\odot}$, but they require host GMCs of masses $10^7 M_{\odot}$ and higher, as noted in Section \ref{sec:mass} above.  The smallest dwarfs will be unlikely to have such massive GMCs and will produce smaller GCs from the start. Subsequent dynamical evolution then plays its role. By the time clusters have evolved to the present day through stellar evolution and tidal dissolution, their masses have been considerably reduced by typical factors of $5 - 10$, and most of the smallest initial clusters have disappeared altogether.  

The dynamical evolution and progressive mass loss of GCs within the potential field of their host galaxy is well-studied, dating back to at least \cite{fall_rees1977} and the definition of the classic GC ``survival triangle'' in the plane of cluster mass and radius. Mechanisms such as tidal shocks, disk shocking, stellar evaporation, and dynamical friction apply for galaxies of all masses, though their relative importance may differ. For example, disk shocking will not be relevant for ellipticals, while tidal stripping from central supermassive black holes will not apply to dwarfs. A few of the huge number of modelling studies that also give extensive references and guides to the literature include, e.g.,  \citet{chernoff_weinberg1990,gnedin_ostriker1997,baumgardt_makino2003,kruijssen2015,webb+2019,krumholtz+2019}. Recent dynamical analyses of dwarf galaxies  include \citet{pennarubia+2009,amorisco2017,contenta+2018,leaman+2020}. As a rough guide, analytical approximations to N-body calculations \citep[e.g.][]{lamers+2005,lamers+2010,webb_vesperini2018} show that to survive for $> 10$ Gy, the initial cluster mass needs to be $\gtrsim 2 \times 10^4 M_{\odot}$ (within factors of a few, depending on the ambient gas density at time of formation, the stellar IMF, and the surrounding tidal field).  These are, however, only general guidelines and the dwarf-centered papers cited above demonstrate the wide range of individual histories that are possible.

\begin{figure}
    \centering
    \includegraphics[width=0.5\textwidth]{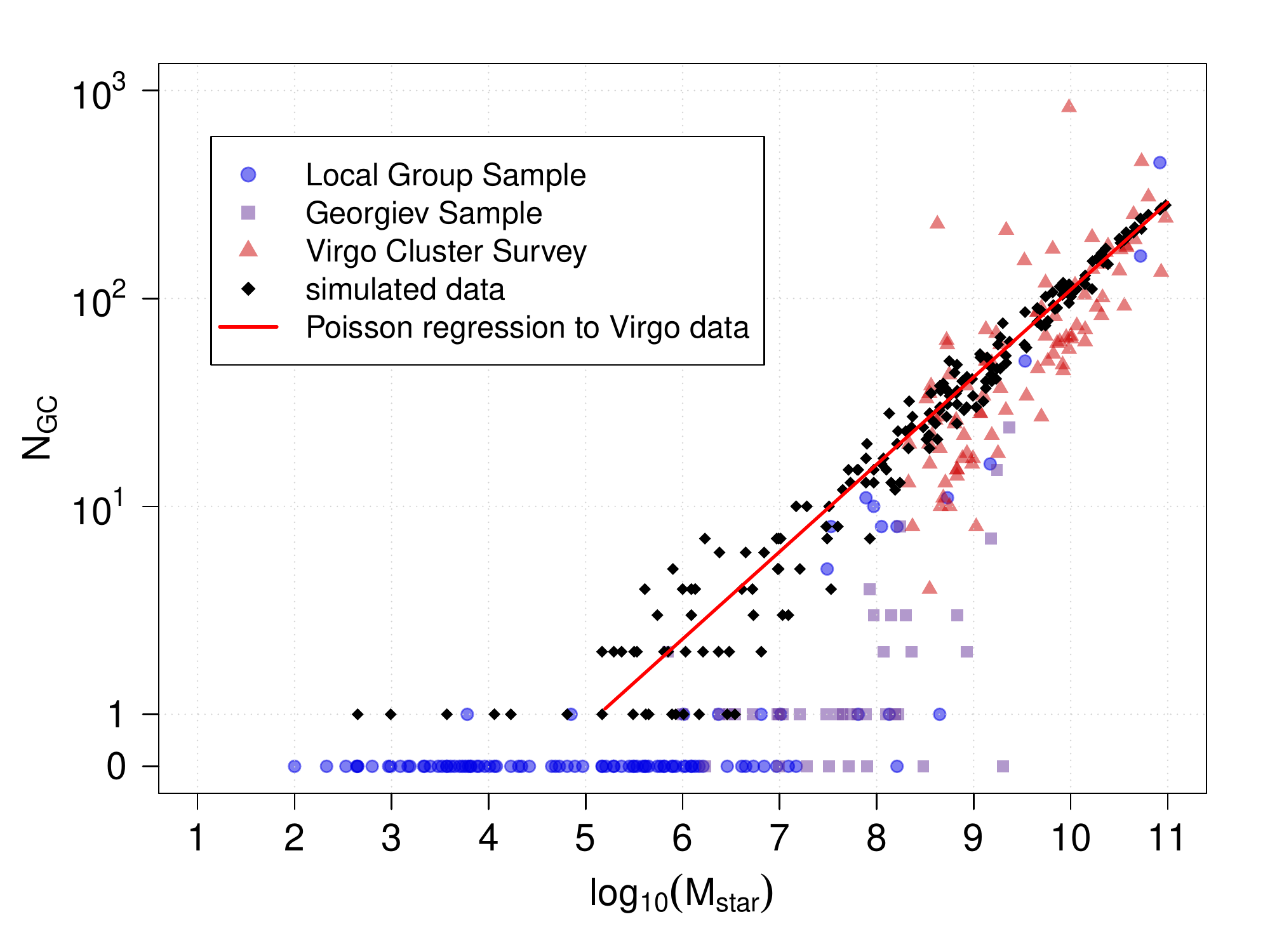}
    \includegraphics[width=0.48\textwidth]{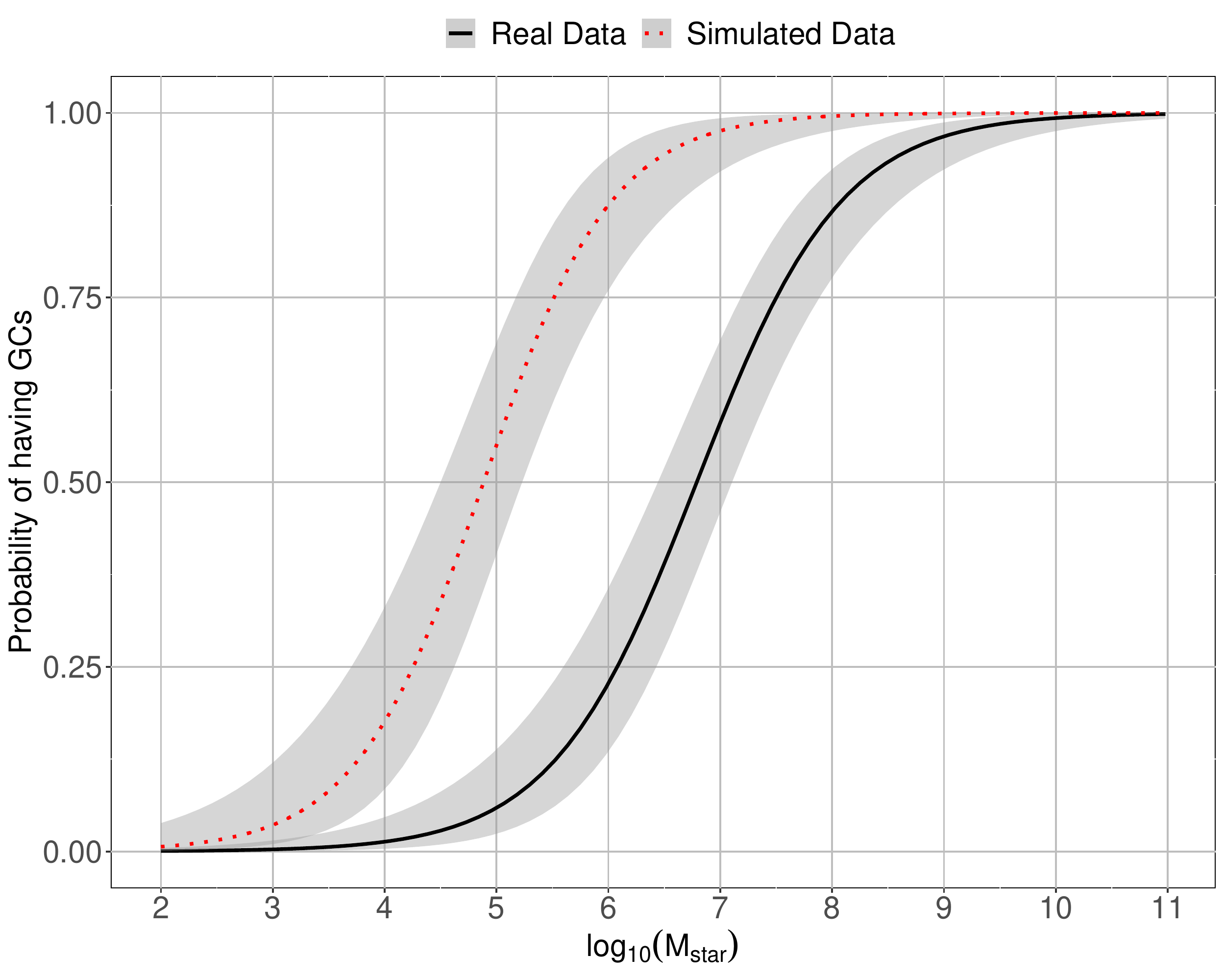}
    \caption{\textit{Top:} The Poisson regression (red line) is fit to the Virgo cluster survey sample, and then simulated data (black diamonds) are generated from this model. \textit{Bottom:} The red dotted curve is the logistic regression fit to the simulated data above (black diamonds), and the solid black curve is the logistic regression fit to the entire set of the real data (Fig.~\ref{fig:logisticregression}).}
    \label{fig:logisticregressionsim}
\end{figure}

Overall, the presence or absence of any GC in a small galaxy at the present day is a complex result of where each GC was originally formed within its galaxy, with what mass and central concentration, and with what velocity. The shape of the dwarf galaxy's DM halo, including the presence or absence of a central cusp, is also a key factor for determining the orbital evolution of the GC and its eventual dissolution \citep{amorisco2017,webb_vesperini2018}.  The range of possible outcomes is large even for the smallest systems
\citep[see, e.g.][for detailed investigations of the possible histories of the GCs in the tiny LG dwarfs EriII, AndXXV, and the Pegasus dIrr]{amorisco2017,contenta+2018,leaman+2020}.

Despite considerable variance among different galaxies, the general picture that emerges is that dwarfs with $M_{\star} < 10^6 M_{\odot}$ will generate star clusters with initial masses of a few $10^4 M_{\odot}$ at most. Dynamical evolution over the following Hubble time will destroy almost all of these clusters, leaving the tiniest galaxies without star clusters now --- even though they were perfectly capable of generating small star clusters initially.
Therefore, we conjecture that the relationship between galaxy host stellar mass and $N_{GC}$ and $M_{gcs}$ (presented in Figures~\ref{fig:dataonly}a, b, and c) is an evolving one: the dwarfs that currently have GCs will gradually lose them through physical processes and evolution, and the transition mass region will gradually shift to higher mass as galaxy and GC evolution continue.

The original goal of this study was to investigate the transition region ($10^5-10^8 M_{\odot}$) between higher-mass galaxies that have GCs and lower-mass ones that do not, and to find an empirical model that can describe this region. Ultimately, we settle on the Bayesian lognormal hurdle model as a way to describe the transition mass region.

In nature, both physical and statistical processes operate simultaneously. Our findings suggest galaxies and their GC systems within the transition region are no different; in this zone, the numbers of GCs are small and subject to large random fluctuations, but at the same time
the individual GCs themselves have low masses and are more likely to be removed through dynamical evolution.

The data analyzed here give us a snapshot of the transition region at the present epoch, but the zone is a continually evolving one.  Ideally, we would like to know about the transition region at a time closer to the GC formation epoch, since this would help us understand GC formation in the context of galaxy evolution.

If we make a simple assumption that
the Poisson model was originally a good description of $N_{GC}$ vs. stellar mass at early times, then we can simulate data that might be more representative of GC counts for dwarfs at early epochs. These simulated data are shown as black diamonds in the top panel of Figure~\ref{fig:logisticregressionsim}. In the bottom panel of Figure~\ref{fig:logisticregressionsim}, we perform a logistic regression on these simulated data, in the same manner as in Section~\ref{sec:moredata}, which is shown as the dotted red line. The logistic regression fit to the real data is shown as the solid black curve, for comparison. Notably, the 50\% probability point for a galaxy to have GCs in the simulated data is $\lesssim 10^5 M_{\odot}$, about two orders of magnitude lower than the 50\% point in the real data. The red dotted line might be closer to the transition region mass range at the epoch of GC formation, when dwarf galaxies as small as $10^5 M_{\odot}$ may have had clusters. Of course, a major caveat here is that we used a Poisson model to generate the simulated data, which we already showed was not an accurate description of the current data. More extended work with simulations of galaxy formation and their accompanying GCs will be needed to explore this direction further.

\section{Summary}\label{sec:summary}

We conclude with a summary of our findings:
\begin{itemize}
    \item We collate a sample of 232 galaxies with well known numbers of globular clusters from the Local Group, nearby relatively isolated dwarfs, and the Virgo cluster.  Using a logistic regression analysis, we estimate the transition region of galaxy mass below which a galaxy is unlikely to hold any GCs today (i.e.~at zero redshift). From our analysis of the LG sample, a galaxy with $M_{\star} = 10^{7.20} M_{\odot}$ is expected to have a 50\% probability of holding at least one GC. This stellar mass corresponds roughly to a halo  (virial) mass $M_h \simeq 10^{10} M_{\odot}$ depending on the adopted SHMR transformation. When using the entire galaxy sample, however, we find the mass corresponding to 50\% probability to be slightly lower, at $M_{\star}=10^{6.79}M_{\odot}$. 
    
    \item The probability of a galaxy hosting GCs is informed by galaxy stellar mass first and foremost. Other predictor variables can be included in a logistic regression analysis; as an example, we explored using approximate values for galaxy morphological type as an extra predictor, or as an extra predictor plus an interaction term with mass, but we found this predictor provided little benefit. Other predictor variables, such as galaxy colour, might be more useful.
    
    \item The transition region is broad, spanning four decades in mass. The point of 10\% probability is reached at $M_{\star} = 10^{5.4} M_{\odot}$ (or $M_h \sim 10^9 M_{\odot}$), and 90\% probability at $M_{\star} = 10^{8.2} M_{\odot}$ ($M_h \sim 4 \times 10^{10} M_{\odot}$).
    
     \item For galaxies that do have GCs, there is a well-defined relationship between the total mass in GCs and the galaxy stellar mass which we first modelled with a linear regression (Section~\ref{sec:LR}): $\log{M_{gcs}} = -0.73 + 0.79 \log{M_{\star}}$ with 95\% confidence intervals on the coefficients of (-1.28, -0.17) and (0.73, 0.85) respectively.  This relation is valid for $M_{\star} < 10^{11} M_{\odot}$ and extends to the smallest known galaxy with a GC, at $M_{\star} \sim 10^5 M_{\odot}$.     Because the stellar-to-halo mass ratio decreases towards less massive galaxies, this finding is consistent with the previously known near-linear proportionality ${M_{gcs}} \sim {M_h}$.
    
    \item We find that an empirical Poisson regression model cannot describe the observed relationship between the number of GCs and host galaxy mass across the range of our data nor for galaxies below $10^8M_{\odot}$. We found the data set to be lacking the appropriate number of dwarf galaxies with $N_{GC}=1,2,3 \text{~or~} 4$, as evidenced by the model prediction intervals, simulated data, checks of the deviance residuals, and results of the dispersion test.
    
    \item We suggest that the pattern of $N_{GC}$ seen in dwarfs with $M_{\star} \lesssim 10^8 M_{\odot}$ is determined by both statistical noise and by GC formation history and subsequent dynamical evolution.  GCs need to have initial masses higher than $\sim 2 \times 10^4 M_{\odot}$ (and perhaps much higher) to have a good probability of surviving for more than a Hubble time, and only dwarfs above $M_{\star} \sim 10^6 M_{\odot}$ were capable of generating such clusters. A wide range of individual histories in the dwarf galaxies appears to be the major reason why the transition region is as broad as it is.
    
    \item The smallest dark-matter halos with virial masses 
    $M_h < 10^9 M_{\odot}$ are extremely unlikely to contain any GCs today, and for $M_h < 10^8 M_{\odot}$ no known galaxies contain GCs. So far, there is no evidence of star-free ``empty'' dark-matter halos with GC systems, but finding any such objects (at any halo mass) would be remarkably important.

    \item A big caveat in this and other recent studies 
    is that in our database, almost all the galaxies that do not have GCs (and that are \emph{known} to have no GCs) are from just the Local Group sample. 
    This limitation no doubt has affected our analysis, and especially the estimation of any interaction term between galaxy type and stellar mass.  
    Additional data for GCs in the faintest
    dwarfs within other environments would considerably help our analysis.
    
    \item We introduced and applied a Bayesian lognormal hurdle model to our data set. A Bayesian lognormal hurdle model analysis of the entire data set provides Bayesian credible regions that are useful in two ways: they can be used to predict (1) the probability of a galaxy having GCs given its stellar mass, and (2) the mass of the GC system around a galaxy given its stellar mass, regardless of whether a GC system is known to exist. The linear portion of the hurdle model provides similar coefficients to the linear regression ($\log{M_{gcs}} = -0.83 + 0.80 \log{M_{\star}}$ with 95\% credible intervals of (-1.41, -0.29) and (0.74, 0.86) respectively), but also provides coefficients for the logistic regression portion ($\beta_0=-10.83 (-14.06,-8.23)$ and $\beta_1=1.59 (1.22,2.05)$).

\end{itemize}

Improvements to our analysis could be made by incorporating a measurement model into the hurdle model (i.e. a hierarchical model). Such a study could include the uncertainty in $\log{M_{\star}}$ associated, for example, with luminosity-to-mass conversion, and this uncertainty would be naturally propagated through to the posterior distribution. On the observational side, surveys of more galaxies, especially in and below the transition region, would be extremely valuable.  Imaging with upcoming wide-field space telescopes such as EUCLID and the Nancy Roman Space Telescope would be superior for searching for small, old star clusters in very low-mass dwarfs both in and beyond the Local Group. As noted previously, currently available GC-count estimates for faint dwarf galaxies beyond the Local Group were mostly excluded from our analysis because their statistical estimates of $N_{GC}$ were quite uncertain. Incorporating these galaxies and their count uncertainties could provide valuable improvements, for example, to the hurdle model fit. Another step forward would be to weight the galaxies in the different samples by the proper proportions of galaxy types and masses in the universe (i.e. matching the sample to the population). Such advancements to our Bayesian lognormal hurdle analysis would not only provide a better estimate of the transition mass region, but also allow for better comparisons to predictions made by simulations.

\section*{Acknowledgements}

GME and WEH acknowledge the financial support of NSERC (Natural Sciences and Engineering
Research Council of Canada).  We thank Ian Roberts
for connecting WEH with the logistic regression model. We also thank the referees for useful comments that improved this paper.  This research has made use of NASA's Astrophysics Data System Bibliographic Services.

\software{R Statistical Software Environment \citep{RSoftware} and the following R packages: \texttt{brms} \citep{brmsBurkner, BurknerRJournal}, \texttt{Cairo} \citep{cairo}, \texttt{ciTools} \citep{ciToolsmanual}, \texttt{dplyr} \citep{dplyrmanual}, \texttt{ggplot2} \citep{wickamggplot2}, \texttt{xtable} \citep{xtable},   \texttt{ragg} \citep{ragg}, \texttt{rstan} \citep{rstan}, \texttt{stargazer} \citep{stargazer}.}

\appendix

\section{The Central Star Cluster in Ursa Major II}\label{app:photometry}

Ursa Major II, at a distance of 34 kpc \citep{dall'ora+2012}, is one of the lowest-luminosity dwarfs in the Local Group at $M_V = -4.2$.  \citet{Zucker+2006} noted that a small clump of stars near its center could be an embedded star cluster, but the cluster has not been studied further and it does not appear in the compilations of \citet{simon2019} and \citet{drlica-wagner+2020}.

The MAST Archive of HST data contains a pair of images of UMaII (from a survey program GO-14734 for Local Group dwarfs) that fortunately includes the cluster.  We downloaded these images and carried out photometry in order to characterize the color-magnitude diagram (CMD) and radial profile of the cluster and its surroundings.  The images, taken with the ACS/WFC camera, are in the filters (F606W, F814W) and have total exposure times of 4661 sec in each.

\begin{figure}[b]
    \includegraphics[width=0.49\textwidth,  trim={0cm 7cm 0cm 2cm}, clip=true]{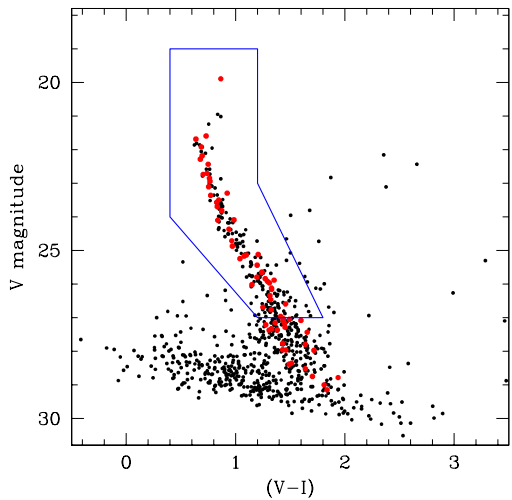}
    \caption{Color-magnitude diagram for the stars measured in the HST/ACS field of Ursa Major II, as described in Appendix~\ref{app:photometry}.  The \emph{red dots} are stars lying within $25''$ of the center of the small central star cluster. The blue polygon encloses an isolated sample of main-sequence stars used for the radial profile analysis (below).}
    \label{fig:umajII_cmd}
\end{figure}

Photometric reduction was done with the standard tools in the \emph{IRAF} versions of \emph{daophot} and \emph{allstar} \citep{stetson1987}, including aperture photometry and PSF fitting.  The \emph{daophot} parameters \emph{chi, sharp} were used to cull out nonstellar objects and thus reduce field contamination.  Conversion of the native F606W, F814W magnitudes to (V, I) was done with the transformations in \citet{saha+2011}. The resulting CMD is shown in Figure \ref{fig:umajII_cmd}.  A well defined main sequence and turnoff region stand out clearly above the field contamination, which is minimal for $V \lesssim 28$.

\begin{figure} 
    \includegraphics[width=0.48\textwidth, trim={0cm 7cm 0cm 1.5cm}, clip=true]{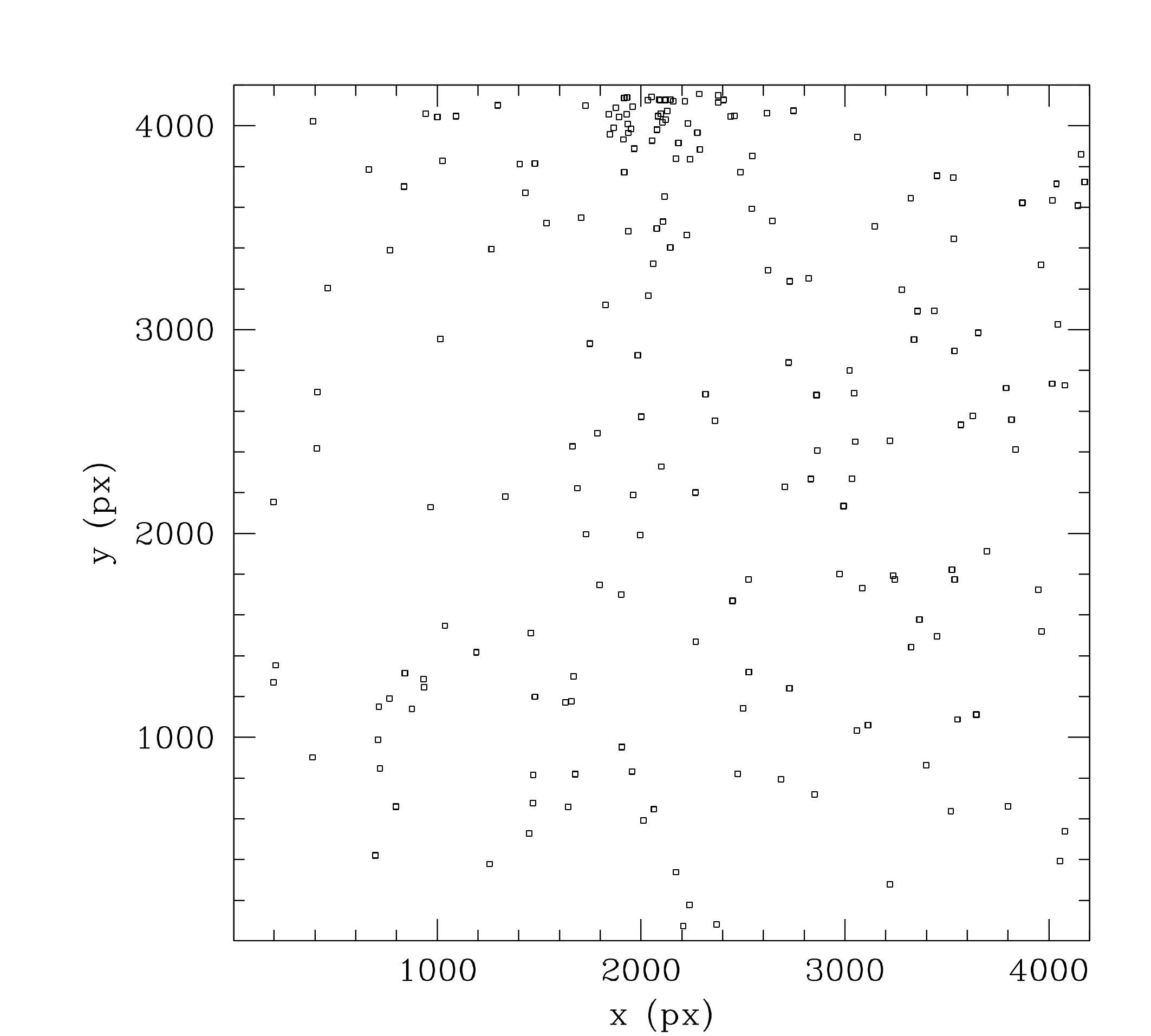}
    \includegraphics[width=0.48\textwidth, trim={0cm 7cm 0cm 1.5cm}, clip=true]{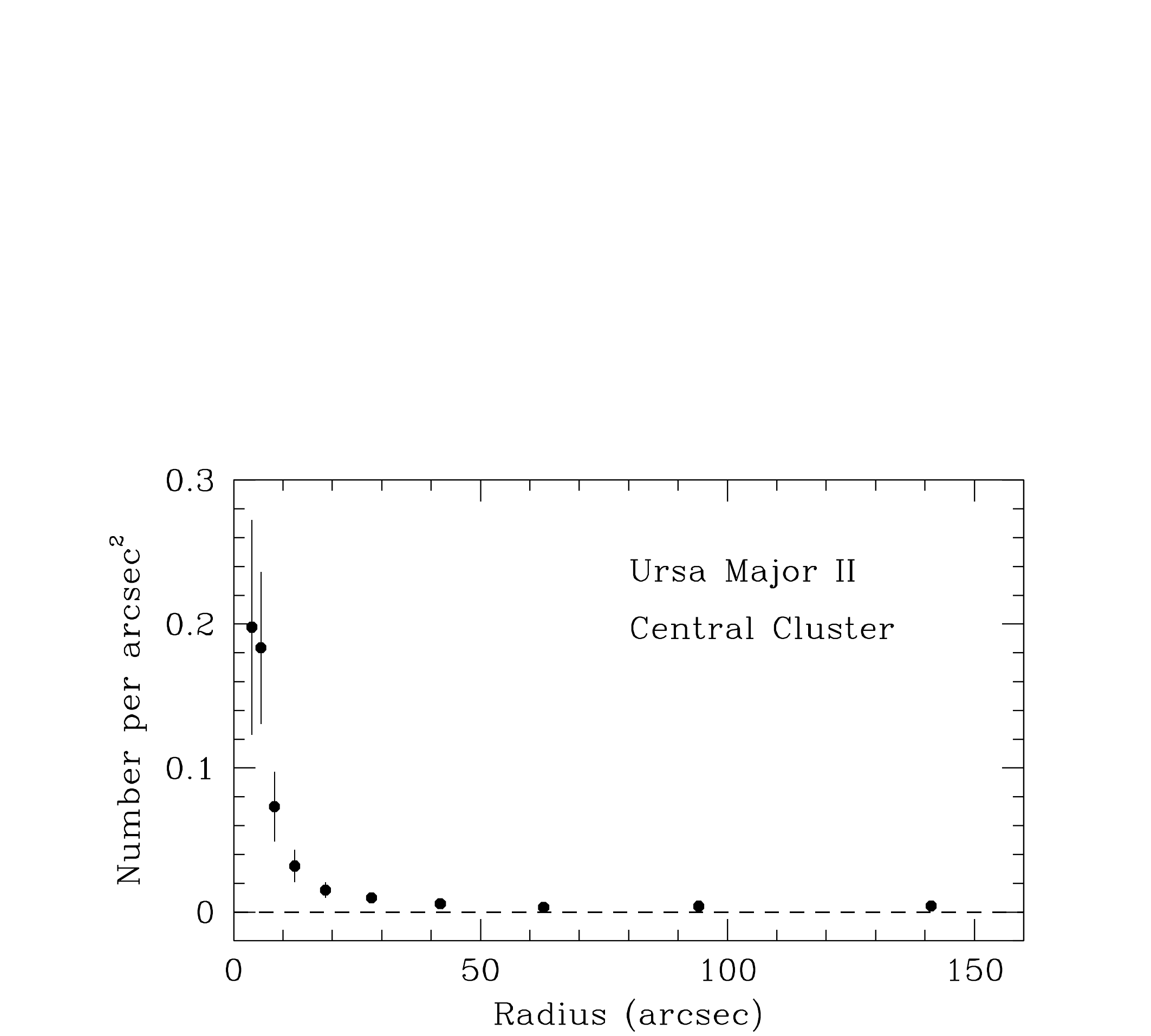}
     \caption{\textit{Left:} Spatial distribution of the main-sequence stars in the UMaII field; these are the stars within the blue polygon from the previous figure.  The star cluster is the small distinct clump near the top edge of the image. \textit{Right:} Number density of stars versus radius from the center of the star cluster.  Stars included here are drawn from the $x,y$ distribution in the previous figure.  Despite its sparse and irregular structure, the cluster shows a well defined profile.  Note the linear scale of the image is 1.6 pc per $10''$.}
    \label{fig:umajII_xy}
\end{figure}

The stars lying within the blue polygon in the CMD were selected to investigate the cluster profile.  These stars are distributed across the image as shown in the left panel of Figure \ref{fig:umajII_xy}.  The cluster shows up as the clump near the top edge of the image, centered at (x,y) = (2010,4040), but it has a somewhat irregular structure with a central `hole' and subclumps of handfuls of stars outside that.  Nevertheless, the azimuthally averaged radial profile
(right panel of Figure \ref{fig:umajII_xy}) is well defined.  The cluster dominates the counts within $r = 25$ arcsec (500 px) of its center; for a distance of 34 kpc \citep{dall'ora+2012}, this corresponds to a linear radius of about 4 pc, well within the range of normal GCs.  The scale size and the close match between the cluster and dwarf stars along the old main sequence (Fig.~\ref{fig:umajII_cmd}) thus support the identification of this object as a globular cluster.  From the CMD, the cluster appears to have a similar age and metallicity to the host dwarf.

For an estimate of the total luminosity of the cluster, we add up all the stars along the main sequence line and within $25''$ of the center, giving $V_{tot} = 18.85$. Using the mean density level of the region outside $r = 25''$ to represent the residual field contamination level indicates that about 7 background stars are present within the cluster circle, or 10\% of the total there.  The contamination-corrected magnitude estimate is then $V_{tot} \simeq 18.95$. The result is, however, especially sensitive to the single brightest star in the CMD at $V = 19.89$, which by itself makes up 40\% of the total cluster light \emph{assuming} that it belongs to the cluster. Finally, for a distance modulus $(m-M)_V = 17.95$ \citep{dall'ora+2012}, the integrated luminosity of the cluster is then $M_V = +1.00$, corresponding to just $34L_{\odot}$.

The cluster mass can be estimated by converting the luminosity $M_V$ of each star to a mass and then summing them up. The PARSEC isochrones \citep{marigo+2017} were used for this purpose, with the assumptions of 12 Gy age and [Fe/H] $= -2$ \citep{dall'ora+2012}.  The resulting total is $52 M_{\odot}$.  This estimate is nominally a lower limit since any main-sequence stars fainter than $V\sim 29$ are not included (though our CMD already indicates that the cluster main sequence becomes very sparse below $V \sim 28$, indicating that its low-mass stars may have been lost during the normal process of dynamical evolution). In addition, unresolved binaries may be present and would increase the mass estimate correspondingly. We tentatively adopt $M_{tot} \simeq 60 M_{\odot}$, recognizing that it is likely to be uncertain by as much as a factor of two given all the issues described above.  The combined evidence -- its irregular structure, very small number of stars, and sparse lower main sequence -- are all consistent with the interpretation  that the cluster is in the very last stages of dissolution.  In the Milky Way, a roughly analogous cluster in a similar late dynamical state would be Palomar 13 \citep{bradford+2011}.

UMaII itself is an extremely sparse object, but it extends across a very much larger area than does the HST/ACS field measured here, and \citet{martin+2008} give $M_* = 5.4 \times 10^3 M_{\odot}$ for it.  The specific mass ratio is then $S_M = 1.1$, though we emphasize this is very uncertain. An alternate interpretation for the UMaII cluster, which is very near the galaxy center, is that it may be the remnant of an NSC (Nuclear Star Cluster).  Measurements of its radial velocity and proper motion would, however, be needed to settle the question.

\clearpage

\section{Hurdle models in R using \texttt{brms}}\label{app:hurdle}

We used the \textit{Bayesian Regression Models using Stan} \texttt{(brms)} package in R to approximate the posterior
distribution of parameters in a Bayesian lognormal hurdle model. An
astronomy-focused discussion of these models is in Section 7.2.4 of \citet{hilbe2017bayesian}.

The \texttt{brms} package provides a domain-specific language (DSL) for
statistical inference using a wide class of models, including the
lognormal hurdle model. Under the hood, the DSL is translated to
\texttt{stan} code and run using \texttt{rstan}. Specifying the
statistical model using \texttt{brms} is similar to specifying other
statistical models in R.

In the first code chunk below, we load the \texttt{rstan} and \texttt{brms} packages, load the data  into the R environment, and look at the first few rows of the data frame. 
\begin{Shaded}
\begin{Highlighting}[]
\FunctionTok{library}\NormalTok{(rstan)}
\FunctionTok{library}\NormalTok{(brms)}

\CommentTok{\# read in data (dataframe object, similar to dataframe in Python), stored in R's serialized format}
\NormalTok{mydata }\OtherTok{\textless{}{-}} \FunctionTok{readRDS}\NormalTok{(}\StringTok{"../Data/best\_sample\_below10to11\_2021{-}01{-}29.rds"}\NormalTok{)}
\FunctionTok{head}\NormalTok{(mydata)}
\end{Highlighting}
\end{Shaded}

\begin{verbatim}
##   GalaxyID Type log10Mgcs elog10Mgcs log10Mstar elog10Mstar Ngc             Sample       Mgcs
## 1     AndI -4.9      3.57        0.2       6.81         0.2   1 Local Group Sample   3715.352
## 2    EriII -3.0      3.60        0.2       4.85         0.2   1 Local Group Sample   3981.072
## 3 Aquarius  9.9      4.16        0.2       6.37         0.2   1 Local Group Sample  14454.398
## 4   AndXXV -3.0      4.18        0.2       6.01         0.2   1 Local Group Sample  15135.612
## 5  PegdIrr  9.8      5.06        0.2       7.01         0.2   1 Local Group Sample 114815.362
## 6      SMC  8.9      5.51        0.2       8.65         0.2   1 Local Group Sample 323593.657
##          Mstar   plotMgcs hasGC
## 1   6456542.29   3715.352     1
## 2     70794.58   3981.072     1
## 3   2344228.82  14454.398     1
## 4   1023292.99  15135.612     1
## 5  10232929.92 114815.362     1
## 6 446683592.15 323593.657     1
\end{verbatim}

The \texttt{brm()} function within the \texttt{brms} package translates a model defined by a formula to \texttt{stan} code, making implementation easier for the user. For example, to generate samples from the posterior distribution using the default priors and sampling control parameters, we pass a \texttt{brmsformula} object to \texttt{brm()}, along with the data, and the type (\texttt{family}) of model to use in a single line of code:

\begin{Shaded}
\begin{Highlighting}[]
\NormalTok{hurdlefit }\OtherTok{=} \FunctionTok{brm}\NormalTok{(}\FunctionTok{brmsformula}\NormalTok{(Mgcs }\SpecialCharTok{\textasciitilde{}}\NormalTok{ log10Mstar, hu }\SpecialCharTok{\textasciitilde{}}\NormalTok{ log10Mstar), }
\AttributeTok{                  family=}\StringTok{"hurdle\_lognormal"}\NormalTok{, }\AttributeTok{data =}\NormalTok{ mydata) }
\end{Highlighting}
\end{Shaded}

The resulting object \texttt{hurdlefit} is of class \texttt{brmsfit} -- for details, see \texttt{?brmsfit}, and \texttt{methods(class\ =\ "brmsfit")} for a list of available methods for posterior analysis.

For those who would like to see what is going on under the hood, the \texttt{make\_stancode()} function allows us to look at the \texttt{stan} code that will be generated by \texttt{brm}:

\begin{Shaded}
\begin{Highlighting}[]
\CommentTok{\# Mgcs \textasciitilde{} log10Mstar is the linear part of the model, and hu \textasciitilde{} log10Mstar is the logistic part}
\CommentTok{\# Check the stan code which will be generated by brm}
\FunctionTok{make\_stancode}\NormalTok{(}\FunctionTok{brmsformula}\NormalTok{(Mgcs }\SpecialCharTok{\textasciitilde{}}\NormalTok{ log10Mstar, hu }\SpecialCharTok{\textasciitilde{}}\NormalTok{ log10Mstar),}
\AttributeTok{              family=}\StringTok{"hurdle\_lognormal"}\NormalTok{, }\AttributeTok{data =}\NormalTok{ mydata)}
\end{Highlighting}
\end{Shaded}

\begin{verbatim}
## // generated with brms 2.14.4
## functions {
##   /* hurdle lognormal log-PDF of a single response 
##    * Args: 
##    *   y: the response value 
##    *   mu: mean parameter of the lognormal distribution 
##    *   sigma: sd parameter of the lognormal distribution
##    *   hu: hurdle probability
##    * Returns:  
##    *   a scalar to be added to the log posterior 
##    */ 
##   real hurdle_lognormal_lpdf(real y, real mu, real sigma, real hu) { 
##     if (y == 0) { 
##       return bernoulli_lpmf(1 | hu); 
##     } else { 
##       return bernoulli_lpmf(0 | hu) +  
##              lognormal_lpdf(y | mu, sigma); 
##     } 
##   }
##   /* hurdle lognormal log-PDF of a single response
##    * logit parameterization of the hurdle part
##    * Args: 
##    *   y: the response value 
##    *   mu: mean parameter of the lognormal distribution 
##    *   sigma: sd parameter of the lognormal distribution
##    *   hu: linear predictor for the hurdle part 
##    * Returns:  
##    *   a scalar to be added to the log posterior 
##    */ 
##   real hurdle_lognormal_logit_lpdf(real y, real mu, real sigma, real hu) { 
##     if (y == 0) { 
##       return bernoulli_logit_lpmf(1 | hu); 
##     } else { 
##       return bernoulli_logit_lpmf(0 | hu) +  
##              lognormal_lpdf(y | mu, sigma); 
##     } 
##   } 
##   // hurdle lognormal log-CCDF and log-CDF functions 
##   real hurdle_lognormal_lccdf(real y, real mu, real sigma, real hu) { 
##     return bernoulli_lpmf(0 | hu) + lognormal_lccdf(y | mu, sigma); 
##   }
##   real hurdle_lognormal_lcdf(real y, real mu, real sigma, real hu) { 
##     return log1m_exp(hurdle_lognormal_lccdf(y | mu, sigma, hu));
##   }
## }
## data {
##   int<lower=1> N;  // total number of observations
##   vector[N] Y;  // response variable
##   int<lower=1> K;  // number of population-level effects
##   matrix[N, K] X;  // population-level design matrix
##   int<lower=1> K_hu;  // number of population-level effects
##   matrix[N, K_hu] X_hu;  // population-level design matrix
##   int prior_only;  // should the likelihood be ignored?
## }
## transformed data {
##   int Kc = K - 1;
##   matrix[N, Kc] Xc;  // centered version of X without an intercept
##   vector[Kc] means_X;  // column means of X before centering
##   int Kc_hu = K_hu - 1;
##   matrix[N, Kc_hu] Xc_hu;  // centered version of X_hu without an intercept
##   vector[Kc_hu] means_X_hu;  // column means of X_hu before centering
##   for (i in 2:K) {
##     means_X[i - 1] = mean(X[, i]);
##     Xc[, i - 1] = X[, i] - means_X[i - 1];
##   }
##   for (i in 2:K_hu) {
##     means_X_hu[i - 1] = mean(X_hu[, i]);
##     Xc_hu[, i - 1] = X_hu[, i] - means_X_hu[i - 1];
##   }
## }
## parameters {
##   vector[Kc] b;  // population-level effects
##   real Intercept;  // temporary intercept for centered predictors
##   real<lower=0> sigma;  // residual SD
##   vector[Kc_hu] b_hu;  // population-level effects
##   real Intercept_hu;  // temporary intercept for centered predictors
## }
## transformed parameters {
## }
## model {
##   // likelihood including all constants
##   if (!prior_only) {
##     // initialize linear predictor term
##     vector[N] mu = Intercept + Xc * b;
##     // initialize linear predictor term
##     vector[N] hu = Intercept_hu + Xc_hu * b_hu;
##     for (n in 1:N) {
##       target += hurdle_lognormal_logit_lpdf(Y[n] | mu[n], sigma, hu[n]);
##     }
##   }
##   // priors including all constants
##   target += student_t_lpdf(Intercept | 3, 12.5, 6.1);
##   target += student_t_lpdf(sigma | 3, 0, 6.1)
##     - 1 * student_t_lccdf(0 | 3, 0, 6.1);
##   target += logistic_lpdf(Intercept_hu | 0, 1);
## }
## generated quantities {
##   // actual population-level intercept
##   real b_Intercept = Intercept - dot_product(means_X, b);
##   // actual population-level intercept
##   real b_hu_Intercept = Intercept_hu - dot_product(means_X_hu, b_hu);
## }
\end{verbatim}
In the example above, the default priors have been used. To set your own prior distribution functions, you can use the \texttt{set\_prior} function and pass this to the ``prior'' argument in \texttt{brm} (see the package documentation for examples). Running \texttt{make\_stancode} is not required for running the \texttt{brm} function; it is meant as a tool for inspecting the stan code that will be generated by \texttt{brm}.

\bibliography{EadieHarrisSpringford2021_2021-10-27_arxiv}
\label{lastpage}

\end{document}